\documentclass[paper,11pt]{article}
\usepackage{graphicx}
\usepackage{bbm}
\usepackage{booktabs}
\usepackage{tabularx}
\usepackage{lscape}
\usepackage{booktabs,caption}
\usepackage{graphicx}
\usepackage[export]{adjustbox}
\usepackage{float}
\usepackage{cite}
\usepackage{hyperref}
\usepackage{epsfig, graphics}
\usepackage{subcaption}
\usepackage[flushleft]{threeparttable}
\usepackage{multirow}
\usepackage{caption}
\usepackage{rotating}

\usepackage{amsfonts}
\usepackage{amssymb}
\usepackage{amsmath}

\newcommand{\be}{\begin{equation}}
\newcommand{\ee}{\end{equation}}

\renewcommand{\thesection}{\arabic{section}}
\newcommand{\ba}{\begin{eqnarray}}
\newcommand{\ea}{\end{eqnarray}}
\newcommand{\ban}{\begin{eqnarray*}}
\newcommand{\ean}{\end{eqnarray*}}

\newcommand{\nin}{\noindent}
\newcommand{\bt}{\begin{tabular}}
\newcommand{\et}{\end{tabular}}

\newcommand{\bge}{\begin{enumerate}}
\newcommand{\ene}{\end{enumerate}}
\newcommand{\bc}{\begin{center}}
\newcommand{\ec}{\end{center}}

\def\bkR{{\rm I\kern-.17em R}}

\def \1n{1\hskip -3pt \mbox{N}}

\def \1n{1\hskip -3pt \mbox{N}}

\newfont{\bbf}{cmbx12 scaled 1435}

\begin{document}
\setlength{\baselineskip}{.26in}
\thispagestyle{empty}
\renewcommand{\thefootnote}{\fnsymbol{footnote}}
\vspace*{0.5cm}
\begin{center}
\setlength{\baselineskip}{.32in}
{\bbf Time-Varying Coefficient DAR Model and Stability Measures for Stablecoin Prices: An Application to Tether}\\

\setlength{\baselineskip}{.32in}

\vspace{0.25in}

\large{Antoine Djogbenou},\footnote[1]{York University, Canada, {\it e-mail}: {\tt
daa@yorku.ca}} \large{Emre Inan},\footnote[2]{York University, Canada, {\it e-mail}: {\tt
emreynan@yorku.ca}} \large{Joann Jasiak}\footnote[3]{York University, Canada, {\it e-mail}:
{\tt jasiakj@yorku.ca}. \\
The authors thank C. Gourieroux and H. Kim and the participants of CMStatistics 2022 and Canadian Economic Association (CEA) 2022 meetings for helpful comments. This project was supported by the Digital Currency Research Clusters Initiative, the Natural Sciences and Engineering Research Council of Canada (NSERC), and the Social Sciences and Humanities Research Council of Canada (SSHRC).
} \\

\vspace{0.25in}

This Version: \today\\

\begin{minipage}[t]{12cm}
\small

\vspace{0.2in}

\bc
Abstract\\
\ec

This paper examines the dynamics of Tether, the stablecoin with the largest market capitalization. We show that the distributional and dynamic properties of Tether/USD rates have been evolving from 2017 to 2021. We use local analysis methods to detect and describe the local  patterns, such as short-lived trends, time-varying volatility and persistence. 
To accommodate these patterns, we consider a time varying parameter Double Autoregressive tvDAR(1) model under the assumption of local stationarity of Tether/USD rates. We estimate the tvDAR model non-parametrically and test hypotheses on the functional parameters.
In the application to Tether, the model provides a good fit and reliable out-of-sample forecasts at short horizons, while being robust to time-varying persistence and  volatility. In addition, the model yields a simple plug-in measure of stability for Tether and other stablecoins for assessing and comparing their stability.

\medskip



\vspace{1in} 

\end{minipage}

\end{center}
\newpage
\setlength{\baselineskip}{.26in}

\renewcommand{\thefootnote}{\arabic{footnote}}
\setcounter{page}{1}

\setcounter{footnote}{0}
\setcounter{section}{0}

\section{Introduction}

The total market capitalization of cryptocurrencies is currently over 1
trillion U.S. dollar, with the top three cryptocurrencies in terms of market capitalization being Bitcoin (BTC), Ethereum (ETH),
and Tether (USDT). While Bitcoin and Euthereum are characterized  by high price volatility, Tether is a stablecoin, i.e. a cryptocurrency 
designed to maintain a stable price compared to other cryptocurrencies such as Bitcoin and Ethereum.
It is the first and by far the largest stablecoin in the market with the highest daily volume of over \$100 billion.
In order to achieve price stability, the value of Tether is pegged 1-to-1 with the U.S. dollar. There also exist other stablecoins with values to
other currency or gold and managed by either a single authority (usually the service provider) or a network of participants (the whole protocol).

Allen, Gu, and Jagtiani (2022) recently discussed how stable cryptocurrencies provide alternative financial instruments for market participants and how appropriately regulated crypto markets could allow increased public confidence and lead to growth and innovation. The November 2021 report by the US President's Working Group on Financial Markets (PWG), the Federal Deposit Insurance Corporation (FDIC), and the Office of the Comptroller of the Currency (OCC) highlight various risks that need to be addressed. These include user protection and run risk, payment system risk, systemic risk and concentration of economic power. They provided various recommendations, including the requirement for stablecoin issuers to be insured depository institutions. See President's Working Group (2021) for more details. Furthermore, Li and Mayer (2021) show that collateralized stablecoins like Tether could create systemic risk if the issuer does not have enough reserve to maintain its stability. More recently, Chen, Qin, and Zhang (2022, page 5) pointed out the important role of Tether in the trading volume of Bitcoin compared to US dollars since 2017 and noted the limited reserve of Tether according to anecdotal evidence.

Despite the increased interest in stablecoins and the recommendations of more scrutiny by regulators, these crypto assets' stability is still ineffective. For example, TerraUSD, an algorithmic stablecoin, collapsed in May 2022. 
This situation posits the need for predictability of stablecoin prices and easy tools for proactive assessment of stability.
	
To address those issues, this paper made the following contributions. First, we analyze the local Tether price from historical data and pin down important features in its dynamic. These features include the local pattern of the mean and the conditional pattern of Tether price as well as the role of specific events in this dynamic. Second, we develop a time-varying model for Tether price that incorporates these specificities. Third, we propose, based on the model, measures that can be used to assess the stability of stablecoins and mitigate risks.

More specifically, we examine the dynamics of Tether/USD rates and documents the time varying distributional properties of this series.
We apply local analysis methods to reveal the time varying mean, volatility and persistence. In particular, we observe periods when Tether rates deviate from the peg, which are often combined with increased volatility.  During those episodes, local persistence measures increase, suggesting unit root dynamics of Tether.

Based on these findings, we consider an extension of the Double-Autoregressive (DAR) model, called the dynamic time-varying parameter Double-Autoregressive (tvDAR).
The DAR model [Ling (2004)] accommodates the conditional heteroscedasticity and nests the ARCH and the autoregressive of order one AR(1) models, including the unit root model with the autoregressive coefficient equal to 1.  More specifically, the DAR is a nonlinear Markov 1 process, which becomes a stationary martingale when the autoregressive coefficient is equal to 1 [Gourieroux, Jasiak (2019)]. The DAR model, unlike the traditional autoregressive AR(1)-ARCH process, provides valid inference and consistent parameter estimators for the autoregressive coefficient values including 1. The proposed extension to a deterministic time-varying parameters model relies on the assumption of local strict stationarity of the process, following the approach of Dahlhaus (2000) and Dahlhaus, Richter, Wu (2019). Then, during the episodes of unit root dynamics, the process satisfies locally the stationary  martingale condition. The time varying tvDAR model provides a good fit to the Tether/USD rates  and gives reliable one step ahead out-of-sample predictions. To obtain the empirical results, we employed a rectangular kernel and an Epanechnikov kernel. The first is an asymmetric kernel, which permits the incorporation of past information in a pre-specified window and could be used for out-of-sample prediction. The second is a symmetric kernel that uses information around any time period and is more suitable for inference on the parameters in the model.

Moreover, the tvDAR model 
provides  a simple  plug-in measure of stability for stablecoins, based on the Lyapunov exponent. This measure is commonly used to assess the stability of deterministic dynamical systems and to test for chaos [see, e.g., Sprott (2014)]. The Lyapunov exponent for the AR(1)-ARCH model has been determined by Borkovec and Kluppenberg (2001), and shown to be the condition of strict stationarity of that process [see also Borkovec (2000)]. It has been also considered by Nelson (1990) in the context of the IGARCH model and by Cline and Pu (2004) in a non-parametric framework. The Lyapunov exponent was also used as a stability measure in application to the Vector Autoregressive VAR model by Dechert and Gencay (1992). Those authors have introduced an alternative stability measure based on the noise-to-signal ratio for linear dynamic models [see also LeBaron (1994) for introduction to chaos].

In this paper, the sample Lyapunov exponent is computed from the model parameter estimates and proposed as a measure of  stability for stablecoins. A more conservative measure, based on the condition of second-order stationarity is also introduced. Both measures can be computed locally and used to assess the stability of a stablecoin over time, or to compare the stability of different stablecoins.  

The time-varying coefficient approach based on the assumption of local stationarity  distinguishes our approach from the literature that relies on the assumption of global strong stationarity of the series. For instance, Baum\"ohl and Vyrost (2022) use high frequency data to compute a spectral density-based quantile dependence measure under a strict stationarity condition, which does not seem to be satisfied by Tether. Bianchi, Rossini, and Iacopini (2022) estimate a Bayesian VAR with stochastic volatility and Student-t distributed shocks (BVAR-SV-t). However, the conditional volatility equation is constrained to unit root dynamics, which is inconsistent with the empirical evidence provided in this paper. 

The paper is organized as follows. Section 2 describes the stablecoins. Section 3 discusses the local dynamic analysis of the price of Tether. Section 4 
discusses the modelling approach, estimation procedures and stability measures. Section 5 presents
the empirical results based on the estimation and inference on the DAR(1) and tvDAR(1) models, including the sample stability measures. Section 6 concludes. Appendix A contains the technical results. Simulation and additional empirical results on stability measures are relegated to Appendices B and C, respectively.


\section{Stablecoins}

This section defines stablecoins and discusses their classification, issuance, and redemption mechanisms. In addition, we discuss how the market prices of stablecoins are determined.

\subsection{Definition and Classification of Stablecoins}

Stablecoins are a type of cryptocurrency designed to maintain a stable price and reduced volatility, compared to other cryptocurrencies such as Bitcoin and Ethereum. Conventionally, stablecoin companies peg the value of their coins to that of a physical asset such as a fiat currency or gold with the assumption that the market price of their coins will eventually stabilize, establishing equivalency with the reference asset. 
The strategies used to achieve price stability of stablecoins are discussed below.

There is currently no standard in place that private enterprises should comply with to qualify as a legitimate stablecoin company. This leaves stablecoin enterprises with unlimited design options to choose from to differentiate their business models. Currently, business models of stablecoin companies differ in  their economic design, the quality of backing they maintain, stability assumptions they rely on, and legal protection they provide for coin holders  (Catalini and de Gortari, 2021). 

While the underlining business models may be diverse and complex, there is interest in the elements of such models to understand their economic implications. For example, one element of interest is the mechanism stabelcoins rely on to stabilize price and another is how the responsibilities are distributed over stablecoin protocols.

There exist two alternative mechanisms used by stablecoin companies  to achieve price stability. They either hold collaterals in their reserves to back the value of their coins or they adjust the supply of coins through software codes to restore the peg with the reference asset. When the market value of a cryptocurrency is backed by collaterals, the cryptocurrency is referred to as a collateralized stablecoin. Conventionally, collateralized stablecoins are split into two sub-categories including off-chain collateralized stablecoins and on-chain collateralized stablecoins. Off-chain collateralized stablecoins are backed by a set of collaterals that have an economic value outside of the blockchain. The reserves of this type of stablecoins usually consist of a fiat currency such as the US dollar for Tether (USDT) or a commodity such as gold for PAX Gold (PAXG). Stablecoins are labelled as on-chain collateralized if the underlying collaterals are composed of crypto assets that are created in a digital form and recorded on a distributed ledger. For instance, Dai (DAI), the largest on-chain stablecoin project, supports 18 collateral assets including not only cryptocurriencies such as Ethereum (ETH) and Chainlink (LINK) but also stablecoins such as Tether (USDT), USD Coin (USDC), TrueUSD (TUSD) and PAX dollar (USDP). 

Some projects opt for developing software codes to minimize price fluctuations instead of collateralizing their coins. This type of cryptocurrencies is called an algorithmic stablecoin as they try to stabilize their price around the peg by contracting or expanding the coin supply with the help of computer algorithms embedded in their design. TerraUSD (UST) was until May 2022 the only example of an algortihmic stablecoin that has a market capitalization over a billion US dollar.

In terms of distribution of responsibilities, stablecoins can be categorized as centralized or decentralized. Centralized stablecoins rely on a single legal entity to maintain the price stability, to manage and protect the collaterals, and to fulfill its obligations to users. For instance, Tether Limited is the legal entity that has the authority as well as the responsibility over every Tether in the circulation. Unlike centralized stablecoins, decentralized stablecoins distribute these responsibilities within their network through smart contracts. This allows network participants to take an active role in determining the rules of the stablecoin protocol such as the set of eligible collaterals and the minimum collateral requirements. The decentralized stablecoin DAI grants users who hold its governance token Maker (MKR) the right to vote on the changes to its protocol.  

Li and Mayer (2021) noted that the introduction of stablecoins is comparable to ``the unregulated creation of safe assets to meet agents' transactional demands'' known as shadow banking. Unlike stablecoin issuers, shadow banks must play the role of credit guarantees in the case of insolvency. However, as we will see later, the observed prices of stablecoins tend to deviate from the peg. In addition, these crypto assets face multiple risks including the risk of liquidation.  

For stablecoins designed to be on par with a fiat currency, the use of reserve allows the stablecoins issuers to sell or buy the currency to achieve its price stability. This mechanism helps stablecoin companies to underpin the market value of their coins and protect against the highly volatile nature of the cryptocurrency markets. It resembles fixed exchange rate regimes currently implemented in Panama, Qatar, and Saudi Arabia. In the fixed exchange rate regime, the central bank also uses its foreign reserves to buy or sell its domestic currency to maintain the fixed parity with the currency peg. When the reserve system fails, the domestic currency can be devaluated. Lyons and Viswanath-Natraj (2020) documented that contrary to central banks with some macroeconomic mandate, including keeping inflation around its target, stablecoin issuers do not have any policy functions. In addition, stablecoin companies cannot use the interest rate or devaluation policy to control the exchange rate. For our analysis, we focus on Tether, which is by far the primarily traded stablecoin in terms of market capitalization. As we will show later, Tether price tends to be noticeably affected by events in the crypto world, leading to deviations from the peg despite using reserves.

Given the aforementioned significant risks for stablecoin holders, there is a need to develop appropriate tools to assess their predictability and stability. This paper develops a model based on the properties of  Tether price and uses it to propose tests for its stability. Before discussing the specificity of the cryptocurrency of interest and the modeling strategy, we provide further explanation on the issuance and redemption of stablecoins.

\subsection{Issuance and Redemption}

Issuance and redemption are the two fundamental market activities that determine the equilibrium quantity of a stablecoin in the market. The equilibrium quantity goes up when new coins are issued, and it goes down when existing coins are redeemed. While the equilibrium quantity changes with issuance and redemption, the price of a stablecoin is held constant during these transactions by the service provider. This constant price policy is the result of the pegging strategy explained in the previous section.

Issuance and redemption of stablecoins are presumably initiated by users.\footnote{See Griffin and Shams (2020) for further discussion on whether Tether issuances are supply-driven or demand-driven.} How these transactions are executed depends on whether stablecoin has a centralized or decentralized structure. 
Issuance takes place following the transfer of funds by user to the stablecoin enterprise. Depending on whether stablecoin is centralized or decentralized, these funds are deposited either into banking accounts of a custodian or into a cryptographical vault. For example, an individual or a business who wants to buy Tether should transfer the funds, specifically the US dollar, to Tether Limited’s accounts at Cathay Bank and Hwatai Bank in Taiwan. The collection of these funds constitutes the reserves of the stablecoin enterprise, and they are meant to be kept as a collateral to back the value of every coin in the circulation. Once the funds are successfully deposited, stablecoins are issued through smart contracts and credited into the user’s wallet. For centralized stablecoins, it is the issuer or the agent that authorizes the issuance of the coins whereas it is done automatically by the blockchain technology for decentralized stablecoins.


Redemption of stablecoins is also initiated by user but the difference is that the transactions take place in the reserve order. To redeem stablecoins, users place an order on the blockchain to exchange their stablecoins for the collateral. Upon the order, the stablecoin enterprise becomes obliged to withdraw the stablecoins from circulation and give the user the corresponding amount of collateral in return. The stablecoins that the user redeems are destroyed subsequently from the protocol. Stablecoin projects pledge in their whitepapers that their coins are 100\% redeemable and redemptions can be performed any time users want at the predetermined price. Hence, one can argue that redeemability becomes the liability of stablecoin enterprises and plays a key role in the sustainability of their projects. It is the issuer that is liable to users for redemptions in centralized stablecoins. Decentralized stablecoins, on the other hand, have no single legal entity that shoulders the responsibility. It is the whole network that is responsible for undertaking redemptions. 

\subsection{Market Price of Stablecoins}

Stablecoin prices are usually fluctuating around the target value. While their value in terms of the reference asset is fixed during issuance and redemptions, they are often traded at a premium or a discount on the exchange platforms.\footnote{See Lyons and Viswanath-Natraj (2020) for the detailed analysis of premium and discount on stablecoin prices.} This splits the market for stablecoins between the primary market and the secondary market, which can be considered analogous to the market for traditional securities. The primary market for stablecoins is where stablecoins are created (issuance) or destroyed (redemption) at the fixed exchange rate predetermined by the stablecoin initiative. The secondary market is where the users trade stablecoins within and across cryptocurrency exchanges such as Binance, Coinbase Exchange, Kraken and Bitfinex. The price of stablecoins in the secondary market is determined as a result of the market activities. According to Griffin and Shams (2020), the secondary market activities account for most of the aggregate Tether flow from 2014 to 2018. Hence, the price of Tether and other stablecoins in the secondary market can be considered as the effective rate at which individuals or businesses can buy and sell stablecoins on a day-to-day basis.

\indent At first glance, the design elements stands out as the primary mechanism through which stablecoin projects try to achieve the price stability. However, the price stabilization in the stablecoin market could be multifaceted. For instance, Lyons and Viswanath-Natraj (2020) argue that the price gap between the primary market and the secondary market, which corresponds to the deviation from the peg, can be mitigated also by arbitrage activities. As long as investors have access to the primary market, the price deviations in the secondary market creates an opportunity for them to make profit. When the price of a stablecoin in the secondary market is above the peg, the arbitrager can buy the coin at the target exchange rate from the primary market and sell it in the secondary market to make profit. This increases the supply of the stablecoin in the secondary market, so it puts downward pressure on its price. Similarly, when the price of a stablecoin in the secondary market is below the parity, an arbitrager can buy the coin from the secondary market and redeem it at the peg ratio in the primary market. In this case, the demand from the arbitrager puts upward pressure on the secondary market price. Hence, one could expect that the price of stablecoins across cryptocurrency exchanges would stabilize around the peg through arbitrage activities. For example, introduction of Tether to the Ethereum blockchain in April 2019, which is associated with increased direct access of investors to the primary market, is found to have a stabilizing effect on the price of Tether in the secondary market (Lyons and Viswanath-Natraj, 2020). 

\indent Bullman et al. (2019) provides the list of alternative tools that each type of stablecoins can adopt to maintain the peg. The list consists of fees, redemption limits, and penalty fees to name a few. Fees and redemption limits could be used by collateralized stablecoins to limit the users' transactions and prevent sudden liquidations while penalty fees can help maintain the minimum level of collateralization. For example, Tether Limited imposes the minimum amount of 100,000 USD required for a fiat withdrawal or deposit and charges the greater of \$1,000 or 0.1\% fee per fiat withdrawal and per fiat deposit.\footnote{https://tether.to/fees/}

The stabilization strategies of Tether have not always been fully successful. The next section presents empirical evidence based on the 
dynamic analysis of the data.

\setcounter{equation}{0}

\section{Tether Dynamics}
This section examines the patterns in Tether dynamics in the sample of $T = 1361$ daily closing prices recorded between November 9, 2017, and July 31, 2021.

\begin{figure}[H]
	\centering
	\includegraphics[width=15cm,height=8cm,angle = 0]{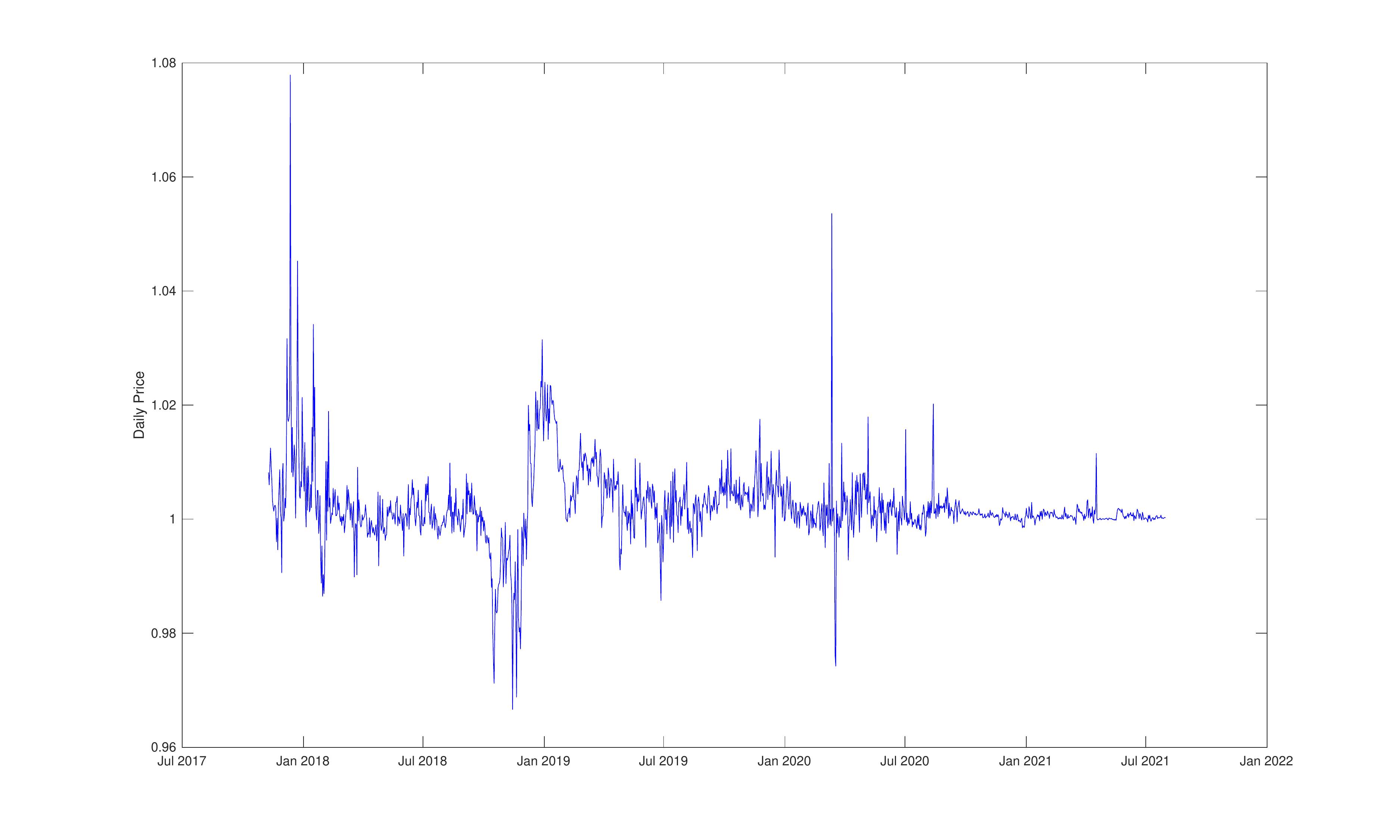}
\caption{Tether/USD daily closing rates}
	\label{localstats}
\end{figure}

\begin{center}
\begin{figure}[H]
\centering
\includegraphics[width=15cm,height=8cm,angle = 0]{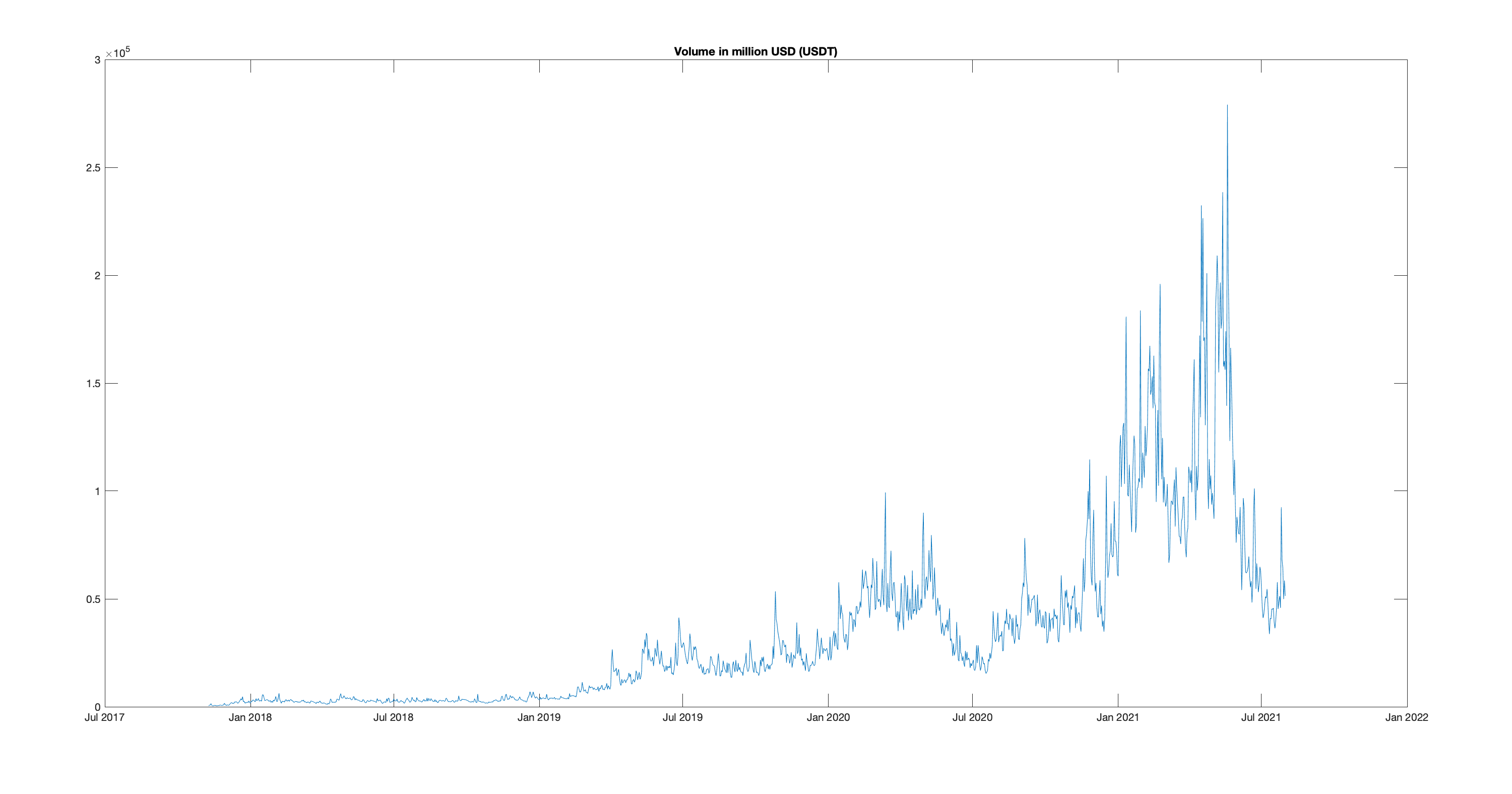}
\caption{Time varying volume of Tether }
\label{close}
\end{figure}
\end{center}

Figure \ref{localstats} displays the evolution of daily closing rates of the Tether against US Dollar. We observe the episodes of explosive dynamics mixed with more stable periods as well as the convergence of the process at the end of the trajectory to a smooth process taking values close to 1. The convergence of Tether towards the peg and its reduced range after 2021 are associated with increased volume displayed in Figure \ref{close}.

\medskip


During the sampling period, the lowest and highest price were 0.9666 and 1.0779, respectively. Although 0.0334 and 0.0779 deviations from the one US dollar parity may look small, they can provide important arbitrage opportunities if the investor is holding a large position in the crypto asset. While the mean over this period is 1.022 and close to one, as expected, the volatility around the mean is 0.066. The evolution of the price shows an alternate of relatively large and small deviation in the stablecoin price due to changes in its demand and lags in intervention by Tether to maintain price stability.

The fluctuation of the Tether price around the one-dollar peg can be connected with the European snake in the tunnel currency system created in April 1972 by an agreement. To increase the convergence among the different currencies in the European Economic Community (EEC), the agreement objective was to create a single currency band within which all the EEC currencies could fluctuate and not deviate too much from a peg. The peg was defined using first gold and, later on, the US dollar. More details on the system can be found in Day (1976). To achieve stability around the peg, central banks had to use their reserve to intervene by buying or selling local currencies. The system was difficult to sustain as several currencies left the agreement. Although stablecoin issuers do not have a macroeconomic policy function as central banks, the difficulties in maintaining the snake currency system also speak to the challenge of maintaining stablecoin prices around its peg using the reserve system discussed above. To understand the price movements of Tether, we first discuss factors that affect its demand during the sampling period.

%

%

The daily volume series in Figure \ref{close} exhibits larger fluctuations during the year 2021 of the sampling period. Although the overall trend was positive, the daily volume varied roughly between \$15.4 billion and \$279 billion. The highest daily volume of \$279 billion was achieved on May 19, 2021, when the Chinese government cracked down on its domestic market for cryptocurrencies. Later, the daily volume plunged to as low as \$33.7 billion in July 2021.

The high levels of daily volume in 2020 and 2021 could be explained by an increasing interest from investors as the market for cryptocurrencies grew substantially during the initial stages of the Covid-19 pandemic.  Tether's daily volume increased drastically from less than \$10 billion in 2018 to as high as \$279 billion in 2021. While the daily volume is observed to be increasing almost steadily in 2018 and 2019, it exhibits rather a volatile pattern in 2020 and 2021. 
For instance, in early 2021, the daily volume of Tether more than doubled in a matter of a few months and reached a peak of 99.3 billion USD on the March 13th, a day after the infamous “Crypto Black Thursday”. However, the pattern in Figure \ref{close} indicates that the daily volume of Tether decreased between May and July of 2021 and returned to its pre-pandemic level. 

The convergence to reduced range and small variation around the constant value of 1 occurs first in Tether in May 2018 and is interrupted by the end of September 2018.  
In the environnement of Tether, the convergence is observed simultaneously  for other mostly traded traded stablecoins such as
USD Coin, Binance USD, True USD, and Pax Dollar
starting from July 2020 and in Dai starting from December 2020. During this period, these 
stablecoins displayed a period of improved stability towards the end of the sampling period.
 Also, the Bitcoin and Etheureum prices in US Dollars have increased. This period of stability overlaps with the period of bullish run in the cryptocurrency market. The cryptocurrency market indices such as the S\&P Cryptocurrency Broad Digital Market (BDM) Index recorded more than fivefold increase between September 2020 and May 2021.\footnote{https://www.spglobal.com/spdji/en/indices/digital-assets/sp-cryptocurrency-broad-digital-market-index/\#overview} The strong demand for cryptocurrencies also benefited the stablecoin companies as the total market capitalization of the top 10 stablecoins went up from approximately \$20 billion in September 2020 to slightly over \$100 billion in July 2021.\footnote{https://www.statista.com/statistics/1255835/stablecoin-market-capitalization/}

The stability is interrupted again for all stablecoins when the cryptocurrency market shrunk by over \$300 billion on April 17, 2021 in less than 24 hours.\footnote{https://www.forbes.com/sites/jonathanponciano/2021/04/18/crypto-flash-crash-wiped-out-300-billion-in-less-than-24-hours-spurring-massive-bitcoin-liquidations/?sh=7d60735b2c89} While the waves of sell-off caused the price of Bitcoin to plummet by 10.5\%, the price of all stablecoins increased simultaneously. This can be evidence in favor of the previous studies which suggest that stablecoins could provide hedging opportunities for cryptocurrency investors against Bitcoin’s volatility, e.g., Wang and Wu (2020). However, one could also argue that the risk mitigating properties of stablecoins, which are closely linked to the comovements between the price of stablecoins and that of Bitcoin, could be changing locally. For example, stablecoins showed resilience against even a much stronger market crash in mid-May 2021. On May 19, 2021, the Chinese government announced that the banks in China are banned from providing cryptocurrency services to their clients.\footnote{https://www.theguardian.com/technology/2021/may/19/bitcoin-falls-30-after-china-crackdown} The market reacted to this news almost immediately as Bitcoin shed 30\% of its value over the course of the day. During the crash, all stablecoins except for Terra USD managed to keep their price stable around the one-dollar peg. 

\subsection{Local Analysis of Tether Price}
This subsection analyzes the local dynamics of Tether price series. We examine its local means, variances, and  autocorrelations. 

\subsubsection{Local mean and variance}

This section studies the evolution of Tether/USD rates and identifies local patterns in  this series by considering time varying descriptive statistics computed by rolling over a window of 50 days.

\begin{figure}[H]
\centering
  \includegraphics[width=15cm,height=10cm,angle = 0]{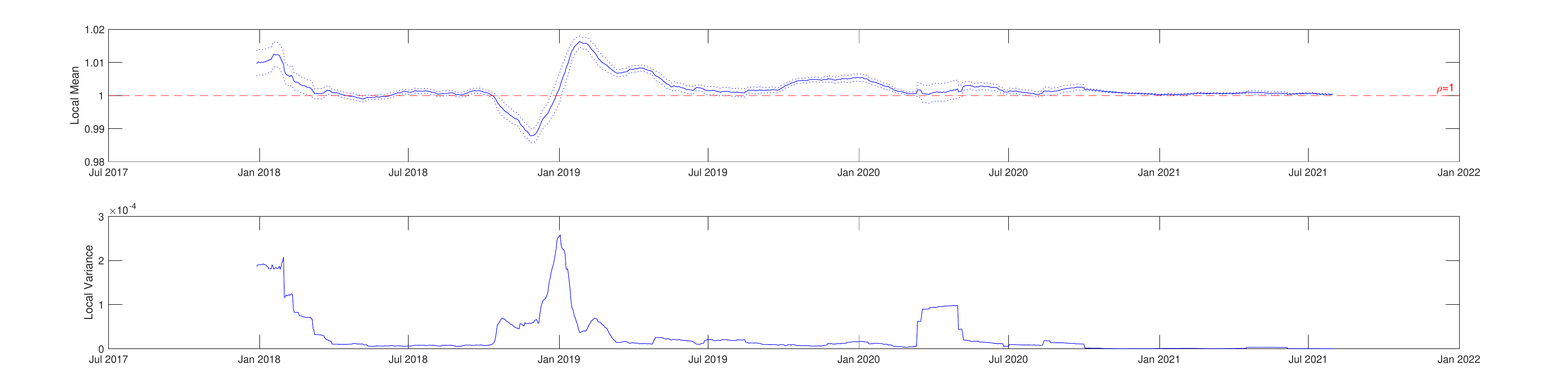}


 \caption{Local mean and variance for the price of Tether}
 \label{TMV}
\end{figure}

The locally estimated marginal mean $\mu_t$ and variance  $\sigma_t^2$  are displayed in panels a) and b) of Figure \ref{TMV}.\footnote{ The lower and upper bounds of the confidence interval at the 95\%  level are calculated under the iid assumption as $\hat\mu_t \mp 1.96 \sqrt{ \frac{\hat\sigma_t}{n}}$ for each window of n days.} 

The figure's top panel reports the local mean of the price series and shows its evolution over the sampling period. We observe that: 
the local mean of Tether varies across sub-periods, and it displays local trends. Especially, in the first half of the sampling period, a strong local trend is observed, which is interrupted by a return of the series to values close to 1. For example, in August 2019, the local mean increases, which is akin to the pattern of financial  bubbles observed in stock prices (rational stochastic bubbles as in Blanchard and Watson (1982) or intrinsic bubbles as in Froot and Obstfeld (1991)). See Kortian (1995) for more details. These patterns, however, disappear towards the end of the sampling period and the local mean becomes more stable and close to the target value of 1. Moreover, there are periods where the target value of 1 falls within the confidence intervals of local means: 
April 19, 2018 to May 5, 2018, May 9, 2018 to June 16, 2018, October 10, 2018 to October 17, 2018, January 2, 2019 to January 3, 2019, and April 2, 2020 to May 1, 2020.
The local variance of the price series is plotted in the bottom panel of the figure. 
It varies over time, and its variation is much higher in the first half of the sampling period. For example, in 2018, Tether had periods of high volatility from January to  March as well as periods of low volatility such as from April to November. On the other hand, Tether is less volatile during the second half of the sampling period  as the local variance takes smaller values except for a short period of increased volatility between mid-March and early May of 2020.

Although the rolling window approach helps detect local trends, it needs to be interpreted with caution. For a window size of n days, the first n-1 observations in the dataset are eliminated due to rolling. In addition, using a longer rolling window (e.g., $100$ days) may over-smooth the changes in the mean and variance as compared to a shorter window (e.g., $50$ days). Therefore, we use 
the window of 50 days for further computations.\footnote{Also, note that n=50 is large enough to estimate parameters within each window consistently. Furthermore, n=50 divided by the sample size of T=1361 is the bandwidth $b_T=n/T=0.0367$ in our local analysis, which will be discussed later. In the literature, an optimal choice should satisfy $Tb_T^3=o(1)$. In our case, we have $Tb^3_T=0.0675$, which is relatively small. See Dahlhaus, Richter, Wu (2019, page 1039) for more details.}

The distributional changes in Tether also concern the range and  quantiles of the series. Overall, we observe that:

\medskip
\nin 1. The local mean is changing over time and is close to 1 between April and June 2018, and after January 2021.

\medskip
\nin 2. The variance is time varying and diminishes over time.

\medskip


In Section 3.2, we identify a series of events that are closely related to Tether, and provide a detailed explanation of the reason why those events could be the driving force behind the changes we observe in the local statistics of Tether.

\subsection{Event Analysis}

The dynamics of Tether are strongly influenced by events, which can be used to distinguish the episodes of distinct patterns in the local mean and variance.

\begin{figure}[H]
\begin{center}
\includegraphics[width=15cm,height=10cm,angle = 0]{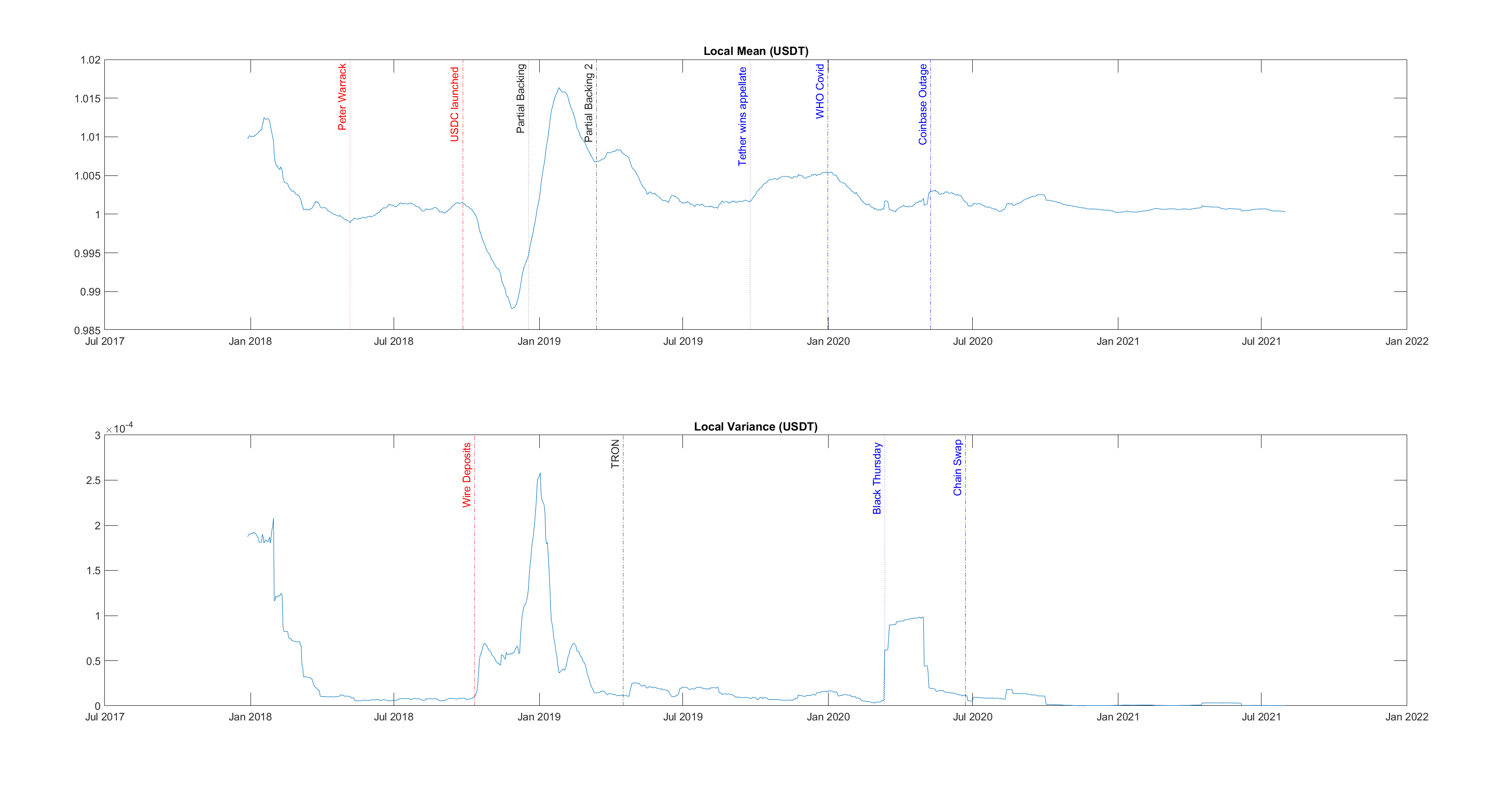}
\caption{Important events for the local mean and the local variance of Tether}
\label{events}
\end{center}
\begin{small}
\textit{Note:} This note provides a description of the events. \\
\nin \textbf{Peter Warrack:} Peter Warrack was hired by Bitfinex as the Chief Compliance Officer on May 7, 2018. \\
\nin \textbf{ USDC launched:} USD Coin was launched on September 26, 2018.\\
\nin\textbf{ Partial Backing:} Bloomberg suggested on December 12, 2018 that Tether could be fully backed.
\\
\nin\textbf{ Partial Backing 2:} On March 14, 2019, Tether made changes to its backing policy on its official website.
\\
\nin\textbf{Tether wins appellate:} Bitfinex won a motion in the New York Supreme Court to delay submission of its business documents.
\\
\nin\textbf{ WHO Covid:} WHO made an announcement on Twitter on December 31, 2019 to acknowledge the cases of pneumonia in Wuhan, China.
\\
\nin\textbf{Coinbase Outage:} Bitcoin shed over 10\% of its value in a matter of minutes on May 9, 2020, which was followed by an outage in the cryptocurrency exchange Coinbase.
\\
\nin\textbf{ Wire Deposits:} Bitfinex ``temporarily paused'' EUR, USD, JPY, and GBP wire deposits on October 11, 2018.
\\
\nin\textbf{TRON:} Tether went live on the Tron network on April 17, 2019.
\\
\nin\textbf{ Black Thursday:} Black Thursday: Bitcoin's price reduced by around 50\% in less than a day on March 12, 2020.
\\
\nin\textbf{ Chain Swap:} On June 22, 2020, Tether announced on Twitter that they would implement a chain swap for a sizable amount of USDT from Tron TRC20 to ERC20 protocol on June 29th.
\end{small}
\end{figure}

\indent Figure \ref{events} shows the evolution of the local mean and the local variance of Tether along with the series of events that can be important for the dynamics of Tether, which can help explain the trend reversals in its local statistics. Total of 11 such events are identified including 7 events for the local mean and 4 events for the local variance.

In 2018, the local mean shows a downward trend for the most part of the year, except for a brief period of recovery between early May and Late September. This should come as no surprise because 2018 was a very tumultuous year for Tether Limited and its business partners. Tether Limited was being scrutinized by the media and scholars for the quality of its reserves and its close ties to the cryptocurrency exchange Bitfinex. More specifically, Tether was publicly accused for not holding enough reserves to back all of its coins in  circulation and for manipulating the price of Bitcoin by pumping unbacked supply of Tether into the market through Bitfinex to buy Bitcoins. Amid these controversies, the local mean of Tether is found to be decreasing for the most part of the year, which could be linked directly or indirectly to a shift in investor sentiment towards Tether.

In 2018, there is also a short period of a slight upward trend in the local mean roughly between early May and late September. In early May, the owners of Tether Limited made their first significant attempt to show their willingness to address the investors` concerns about the accountability of their business and that of Bitfinex. On May 7, 2018, the cryptocurrency exchange Bitfinex officially announced that Peter Warrack, who worked previously at RBC Royal Bank for 20 years as an anti-money laundering specialist, joins their team as the Chief Compliance Officer. Upon this news, the local mean of Tether enjoys a period of recovery and hovers above the one-dollar peg. Nevertheless, the local mean of Tether starts to decrease once again in September and reaches its lowest level in November. The downfall of Tether during this period could be triggered by the introduction of USD Coin (USDC) on September 26, 2018. USD Coin, which is also designed to be on par with the US dollar, relies on the business principles similar to that of Tether but it claims to offer its users an improved transparency in its business activities.

Tether makes up for the loses quickly as the local mean increases remarkably from its lowest level in November 2018 to its highest level in January 2019 just under two months. The surge in the mean price of Tether could be a sign of positive reaction from the investors as the bank statements obtained by Bloomberg showed that on the contrary to the allegations, Tether Limited could be holding enough reserves to back its coins in the circulation.\footnote{https://www.bloomberg.com/news/articles/2018-12-18/crypto-mystery-clues-suggest-tether-has-the-billions-it-promised} However, this positive attitude towards Tether does not seem to last long as the local mean diminishes once again starting in January 2019. The downward pressure on the local mean was so strong during this period that it persisted up until August 2019 despite Tether's effort to be more transparent about the composition of its reserves. According to coindesk.com,\footnote{https://www.coindesk.com/markets/2019/03/14/tether-says-its-usdt-stablecoin-may-not-be-backed-by-fiat-alone/} Tether changed the terms on its website in Mid-February implying that while every Tether is always 100\% backed by their reserves, the reserves are not necessarily composed of only fiat currency as they claimed before but also includes cash equivalents, other assets, and receivables from loans. This update, however, did not stop the New York State Attorney General (NYSAG) sue Bitfinex and Tether Limited on April 24, 2019 on the basis of an ongoing fraud.\footnote{https://ag.ny.gov/press-release/2019/attorney-general-james-announces-court-order-against-crypto-currency-company}

As the downward trend in the local mean dies out in August 2019, another episode of increasing local trend is observed subsequently. The trend becomes more noticeable around September 24, 2019 when iFinex Inc, the parent company of Bitfinex and Tether Limited, won a motion in the court against NYSAG meaning that the company does not have to hand over the documents related to its business activities until further notice.\footnote{https://www.forbes.com/sites/michaeldelcastillo/2019/09/24/bitfinex-and-tether-win-appeal-from-new-york-supreme-court-in-900-million-case/?sh=7c8bf41132bc} Following this small victory in the court, the local mean of Tether diverges from the peg and keeps increasing until the end of the year.

On December 31, 2019, the World Health Organization (WHO) announced via its official Twitter account that they were informed of cases of pneumonia of unknown cause in Wuhan City, China.\footnote{see https://twitter.com/who/status/1213795226072109058?lang=en for the original tweet from WHO} With this announcement, WHO acknowledged the problem of an epidemic in China, which would soon turn into a global health and economic crisis of COVID-19 pandemic. Subsequently, the local mean of Tether starts to decline cancelling out the gains from the last quarter of 2019. 

\subsection{Persistence in Tether Series}

\begin{figure}
\centering
\includegraphics[width=15cm,height=8cm,angle = 0]{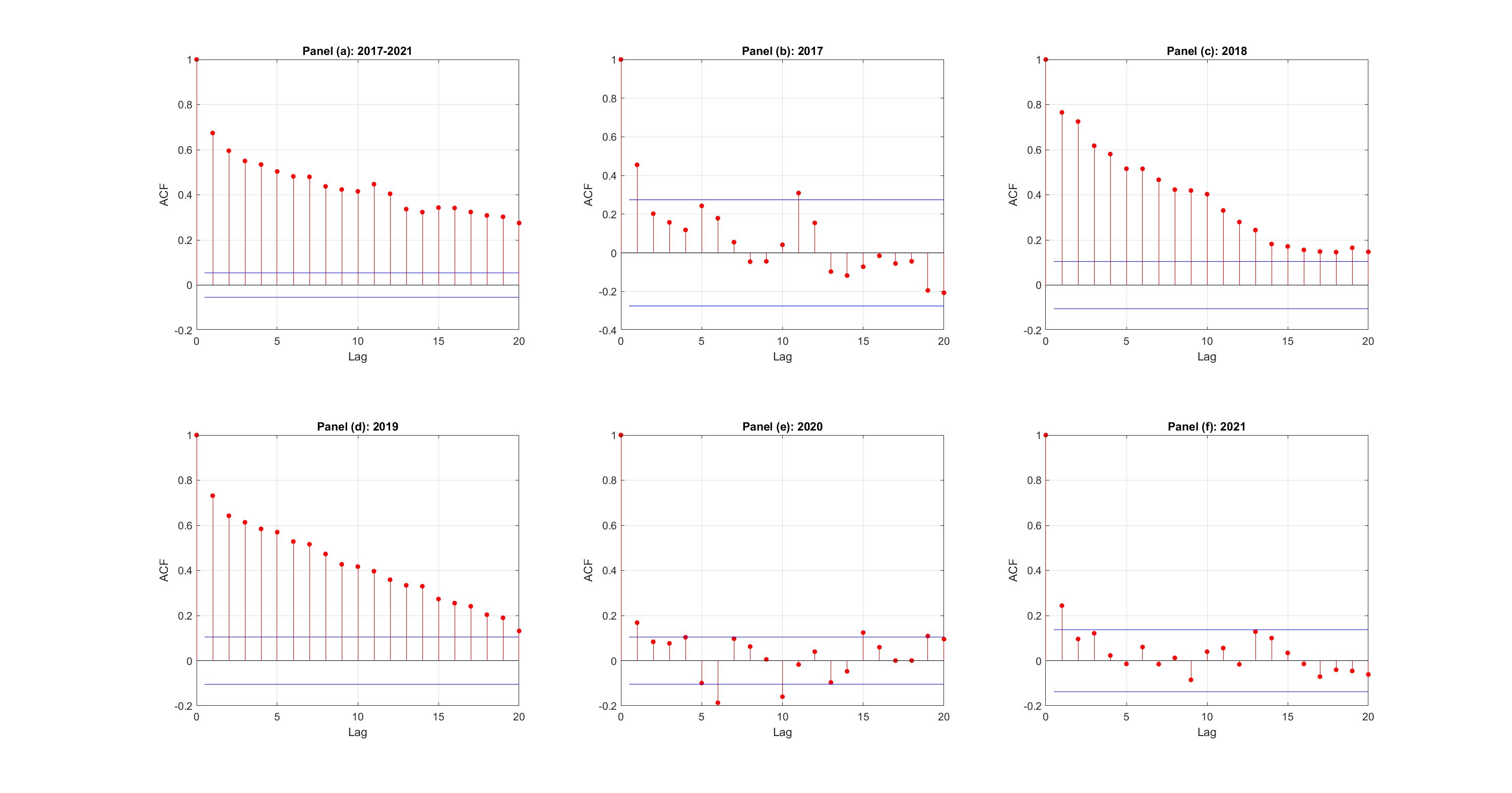}
\caption{Autocorrelation functions of Tether over different periods}
\label{acfs}
\end{figure}

\indent 
Let us now examine the serial correlation in Tether. The autocorrelation function (ACF) of Tether is computed from the entire sampling period 2017-2021 as well as from each calendar year separately. Figure \ref{acfs} presents the computed ACF functions: the ACF over the entire period (panel (a)), and in years 2017-2021 in panels (b) to (e), consecutively.

The ACF calculated from the entire sample exhibits a long range persistence. However, the subperiod analysis reveals that the persistence in the ACF of  Tether is strong up to and including 2019,\footnote{In 2017, there are only 53 observations, which could be the reason for the weak evidence for the persistence.} whereas the series has a short memory in 2020 and 2021.

\indent  When combined with the results from the local statistics, it can be inferred  that the period of long-range persistence in Tether coincides with the period of level shifts and high volatility as documented in Figure \ref{localstats}. Likewise, when the variation is small and the local mean stabilizes around the one-dollar peg as in 2020 and 2021, Tether displays a short memory.


In brief, the empirical results show that the analysis of Tether based on global statistics would provide unreliable results, especially concerning the serial correlation of the series. For example, different  values and range of serial correlation are obtained in year 2021, as compared to years 2018-2019. 

Let us now focus on the autocorrelation values. More specifically,
the autocorrelation at lag one of the series can be estimated from the autoregressive coefficient of an autoregressive of order 1 (AR(1)) model.
We first consider the autoregressive coefficient estimated from the AR(1) model fitted to the whole sample of demeaned Tether prices $x_t=y_t-\mu$ 

\begin{equation}
x_t = \rho x_{t-1} + \sigma_e e_t,
\label{Model1}
\end{equation}

\nin where $e_t$ is assumed to be a white noise with mean 0 and variance 1. The AR(1) process is stationary when $|\rho| <1$ and nonstationary and explosive when $\rho=1$. Model (\ref{Model1}) is 
estimated globally by the OLS, or equivalently by maximizing the Gaussian Maximum Likelihood as follows:
{\bf
$$\hat{\theta}_T = Argmax_{\theta} \sum_{t=1}^T l(x_t| x_{t-1}; \theta) = Argmax_{\theta} \sum_{t=2}^T -\frac{1}{2} \left (\log\left(2 \pi \sigma_e^2\right) + \frac{\left(x_t-\rho x_{t-1}\right)^2}{\sigma_e^2}\right) \, $$}

\nin where $\theta= ( \rho, \sigma_e^2)$. 

\begin{figure}[h]
\centering
  \includegraphics[width=15cm,height=10cm,angle = 0]{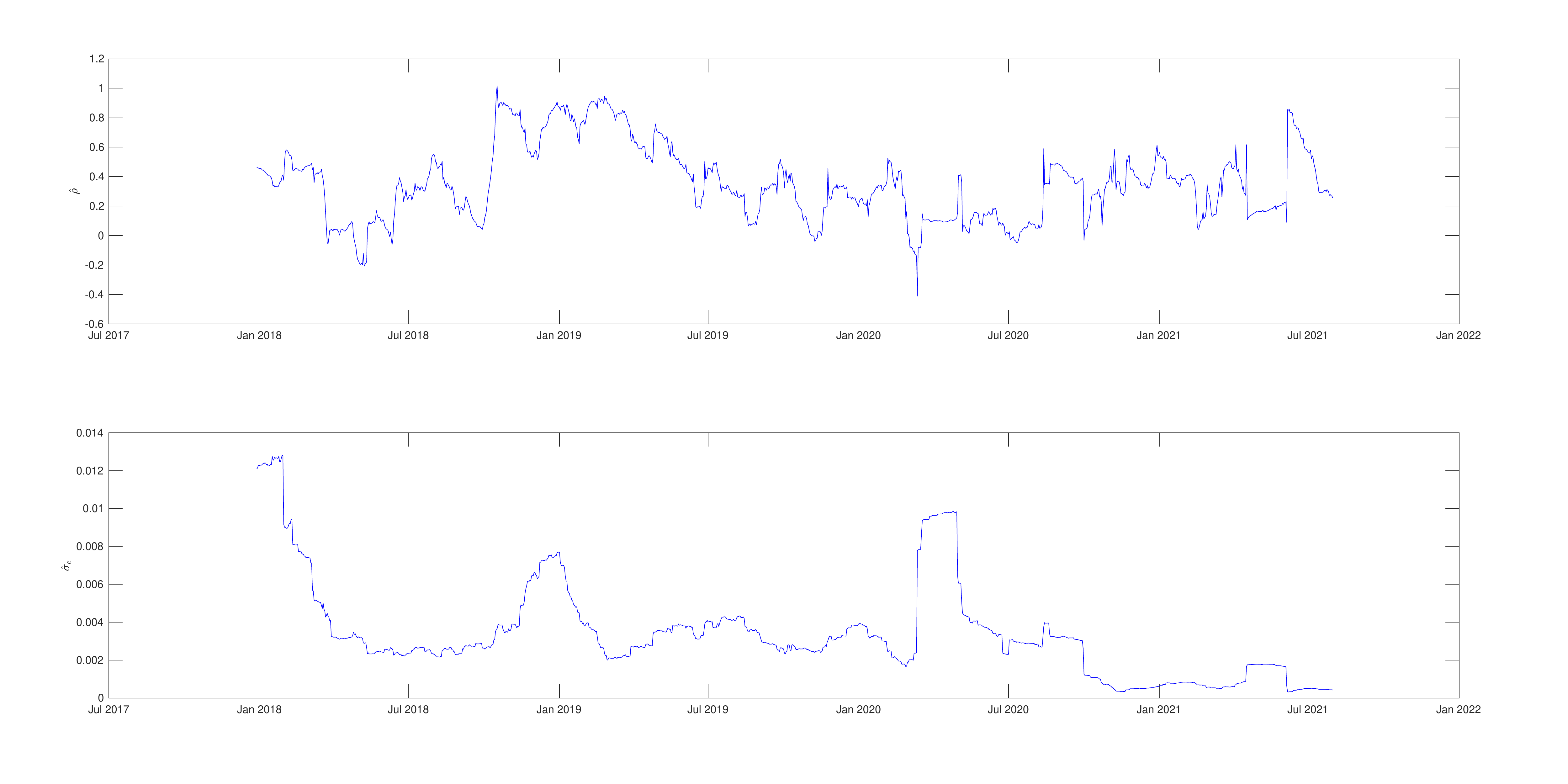}
\caption{AR(1) parameter estimates and conditional volatility for $x_t$}
 \label{AR1}
  \end{figure}

The estimated parameter values are $\hat\rho_T = 0.674$ and $\hat \sigma_{e,T}^2=0.545$. Next, model (\ref{Model1}) is
fitted locally and estimated by rolling with a window of length 50 and displayed in Figure \ref{AR1}.

\medskip

\medskip
The top panel of Figure \ref{AR1} shows the autoregressive coefficient/correlation at lag 1 estimates. We observe that the autoregressive coefficient is close to 1 during the explosive episodes in 2018 and 2019, which violates the stationary condition of the AR(1) process. The unit root dynamics of Tether resembles the stock prices and exchange rates. It suggests that Tether is then locally efficient in financial terms. The autocorrelation at lag 1 is close to 0.5 at the end of the sampling period. 
In comparison with the results presented in Section 3.1, estimating the autocorrelation at lag one based on the AR(1) model gives us greater flexibility to assess the change in the persistence of Tether as we are able to examine its evolution at a daily frequency rather than on a yearly basis. For example, the estimated values of the AR(1) coefficient $\rho$ suggest that the autocorrelation at lag one ranges between -0.20 and 1.02 in 2018 whereas in Section 3.1, the ACF at lag 1 was estimated to be slightly over 0.76 for the entire year. 

The bottom panel shows the conditional volatility $\hat \sigma_e(t) =\sqrt{\frac{1}{50}\sum_{\tau=t-49}^t(x_{\tau}-\hat\rho(\tau)\, x_{\tau-1})^2}$ of the price of Tether under the AR(1) assumption. The figure shows that this price exhibits periods of low volatility and high volatility. Especially from October 2020 onwards, the conditional volatility decreases remarkably and becomes very close to 0. This result is consistent with the period of stability we observed in the price series of Tether during the cryptocurrency bull market explained in Section 3. If the volatility was computed over the full sample we would have a constant estimate which does not capture the changes in volatility over time. To accommodate this feature, we consider the time-varying volatility model which also allows us to have valid inference when the estimated correlation parameter is close to one unlike the AR(1) model.

%

\subsection{Properties of rolling estimators}
 Let us now examine the properties of the rolling estimators of time varying parameters written as deterministic 
functions of time. First, we estimated by rolling the time varying marginal mean and variance functions, hoping to approximate $m(t)$ and $\sigma^2(t)$ in a simple model 
$$y_t = m(t)  + \sigma(t) u_t,
$$
under the simplifying assumption of Normally distributed i.i.d. process $u_t$ with mean 0 and variance 1. Then, in finite sample 

$$\hat{m}(t) = \frac{1}{50} \sum_{\tau=t-49}^t y_{\tau} \sim N \left( \frac{m(t) +\cdots + m(t-49)}{50},
\frac{\sigma^2(t) +\cdots + \sigma^2(t-49)}{250}   \right) $$ 

\nin We see that $\hat{m}(t)$ is biased of m(t) towards an integrated mean. By the same argument it can be shown that $\hat{\sigma}^2(t)$ is biased of both $\sigma^2(t)$ and the integrated variance.

Let us now consider the time varying parameters $\theta(t) = (\rho(t),\sigma^2_e(t)) $ of  a time varying parameter AR(1) model:

$$
x_t = \rho(t) x_{t-1} + \sigma_e(t) e_t,
$$
 
\medskip
\nin where $x_t=y_t-m(t)$. Then, the normality-based MLE estimator $\hat{\theta}^*(t)$ obtained by rolling can be written as the following kernel MLE estimator (see Fan et al. (1998) for the method of local kernel-weighted likelihood estimation using local polynomial fitting).

\begin{eqnarray*}
\hat{\theta}^*_T(t) & = & Argmax_{\theta} \frac{1}{50} \sum_{\tau= t-49}^t \,  l(x_{\tau}| x_{\tau-1} ; \theta) \\
& = & Argmax_{\theta} \frac{1}{50} \sum_{\tau=1}^T 1_{t-49 \leq \tau \leq t} \,  l(x_{\tau}| x_{\tau-1} ; \theta) \\
& = & Argmax_{\theta} \sum_{\tau=1}^T \left[ \frac{1}{50} 1_{-49 \leq \tau-t \leq 0} \,  l(x_{\tau}| x_{\tau-1} ; \theta)  \right] \\
& = & Argmax_{\theta} \sum_{\tau=1}^T \frac{1}{50} \,  1_{-49/50 \leq (\tau-t)/50 \leq 0} \, l(x_{\tau}| x_{\tau-1} ; \theta) \\
& = & Argmax_{\theta} \sum_{\tau=1}^T \frac{1}{50} \, K \left(\frac{\tau-t}{50} \right) \, l(x_{\tau}| x_{\tau-1} ; \theta), \; t=1,...,T
\end{eqnarray*}

\nin with the kernel $K(u) = 1_{[-1,0]}(u)$.

\nin It is easy to see that the dimension of the parameter of interest $[\theta(1),...,\theta(T)]$ depends on the number of observations T. To circumvent this difficulty, we can replace the functional parameter $[\theta^*(1),...,\theta^*(T)]$
with $t \in N$ by an alternative functional parameter $\theta(c), c \in (0,1)$ on [0,1], such that $\theta^*(t) = \theta(t/T)$. The rolling MLE is such that

$$
\hat{\theta}_T^*(t) = \hat{\theta}_T(t/T) =  Argmax_{\theta} \sum_{\tau=1}^T \frac{T}{50} \, K \left(\frac{\tau/T-t/T}{50/T} \right) \, l(x_{\tau}| x_{\tau-1} ; \theta), \; t=1,...,T. $$

\nin This formula can be extended to any value of argument $c \in [0,1]$:
$$ 
\hat{\theta}_T(c) =  Argmax_{\theta} \sum_{\tau=1}^T \frac{T}{50} \, K \left(\frac{\tau/T-c}{50/T} \right) \, l(x_{\tau}| x_{\tau-1} ; \theta), \; c \in [0,1].$$

\nin Dahlhaus (2000) and Dahlhaus, Richter, Wu (2019) show that under regularity conditions the functional parameters $\theta(c)$ of a locally stationary process can be consistently estimated. Instead of considering the time varying parameters in calendar time, we have now defined the functional parameters in a deformed time
$ t \rightarrow t/T$ that depends on T. The functional parameter is now independent of the observations, while the effect of T is introduced by considering a triangular array approach. This leads to a sequence of models indexed by T:

$$ x_{t,T} = \rho(t/T) x_{t-1,T} + \sigma_e (t/T) e_{t,T},$$
where  $x_{t,T}=y_{t,T} - m(t/T)$.

\nin This approach motivates our modelling approach presented in the next section.

\section{DAR(1) Model for Tether Price}

To account for time-varying conditional mean and volatility, we introduce the  Double Autoregressive (tvDAR) process of order 1 with time varying parameters.  The first part of this section recalls the constant parameter DAR model. Next, the estimation of both type of models is discussed. The last part of this section presents the stability measures and their estimators.

\subsection{Model with Constant Parameters}

We consider the DAR process of order 1 for the demeaned Tether price series:

\begin{equation}
x_t = \phi x_{t-1} + \eta_t \sqrt{\omega + \alpha x_{t-1}^2},
\label{DAR10}
\end{equation}

\nin where $\omega>0, \alpha >0$, and $\eta_t, t=1,...,T$ is an independent and identically distributed (i.i.d.) sequence with mean 0 and variance 1. The parameter $\phi$ captures 
the conditional mean dependence. Parameter $\alpha$ represents the past dependence in the conditional variance. The model is semi-parametric and conditionally heteroskedastic.\footnote{More on the models with conditional heteroscedasticity and their applications in finance can be found in Gourieroux (1997). Zakoian (1994) also proposed maximum likelihood and least squares estimators for conditionally heteroscedastic model with threshold.}  Borkovec and Kluppenberg (2001), Ling (2004) show that there exists a unique strictly stationary and ergodic solution to model (\ref{DAR10}) when the following assumptions hold:

\medskip

\nin \textbf{Assumption A.1:} $\eta_t$ has a symmetric and continuous density with mean 0 and variance 1.

\nin \textbf{Assumption A.2:} The parameter space is $\Theta = \{ \theta = (\phi, \omega, \alpha):  E(ln|\phi + \eta_t \sqrt{\alpha}|) <0$ with
$ |\phi| \leq \tilde{\phi}, \underline{\omega} \leq \omega \leq \tilde{\omega}, 
\underline{\alpha} \leq \alpha \leq \tilde{\alpha}$ where $\tilde{\phi}, \underline{\omega}, \tilde{\omega},\underline{\alpha}, \tilde{\alpha}$ are positive constants.

\medskip

 Assumption A.1 is not a stringent assumption. It is satisfied in particular if $\eta_t, t=1,...,T$ are normally distributed. Assumption A.2 is the existence and negativity of the Lyapunov exponent ensuring the existence and uniqueness of a stationary solution [Borkovec and Kluppenberg (2001)].
The region of $\phi, \alpha$ that satisfy the negativity condition is displayed in Figure 1, p. 64 Ling (2004) and Figure 1, p. 191 Chen et al. (2014). It includes cases when $\phi \geq 1$ as well as $E(x_t^2)=\infty$. The processes $x_t$ that satisfy Assumption 2 are strictly stationary. Some of these processes are also weakly (second-order) stationary and satisfy additionally
the 
condition $\phi^2 + \alpha <1$ ensuring that $E(x_t^2) < \infty$.  Thus, the marginal variance of those processes is finite. 
 
When $\phi =1$, and
$E(\ln|1 + \eta_t \sqrt{\alpha}|) <0$, the process $x_t$ is a strictly stationary martingale process
with volatility induced ``mean-reversion'' [Gourieroux, Jasiak (2019)].  Model (\ref{DAR10}) is non-stationary
when the Lyapunov exponent is non-negative. In particular, it is nonstationary at the boundary points $(\phi, \alpha) = (\pm1,0)$ and nests the standard unit root models at these two points. 
When $\phi=0$, the process is an ARCH(1) model. Moreover, process $(x_t)$ is strictly stationary when $x_0$ is drawn from a stationary distribution.

Under assumptions A.1 and  A.2, the parameter space $\Theta$ is compact and there exists a unique strictly stationary solution of the model for any $\theta \in \Theta$. In addition, we assume that:

\nin \textbf{Assumption A.3:} The model is well-specified, i.e. the process satisfies equation (4.2) for the true value of parameter $\theta_0 = (\phi_0, w_0, \alpha_0)$ and the true density $\psi_0$ of 
$\eta$. The true parameter value $\theta_0$ is an interior point in $\Theta$.

\nin \textbf{Assumption A.4:} The observed process is the unique, strictly stationary solution associated with $(\theta_0, \psi_0)$. 

\nin These two conditions are introduced for the identification of the model and parameter estimation.

\subsection{Model with Time-Varying Parameters}

The DAR(1) model can be extended to a time-varying parameter model by using the triangular array approach for locally stationary processes [Dahlhaus (2000), Dahlhaus, Richter, Wu (2019)]. The time varying tvDAR(1) model is written
for locally demeaned observations $x_{t,T}, t=1,...,T$ indexed by $t$ and $T$ (triangular array) and defined by:

\begin{equation}
x_{t,T} = \phi(t/T) x_{t-1,T} + \eta_{t,T} \sqrt{\omega(t/T) + \alpha(t/T) x_{t-1,T}^2},
\label{DARloc}
\end{equation}

\nin where for each time $T$, $(\eta_{t,T})$ is a strong (i.i.d) white noise with mean zero, unit variance and a symmetric distribution invariant in $T$. $\phi(c), \omega(c)>0, \alpha(c)>0$, $c \in [0,1]$ are deterministic functions. We assume that these functions are smooth.

\medskip

\nin \textbf{Assumption a.1:} The functions $\phi(.), \omega(.), \alpha(.)$,  are positive, deterministic and twice differentiable on $[0,1]$.

\medskip

Moreover, the trajectories of the process have to be little responsive to small changes of the parameters, which is ensured by  a Lipschitz condition. More precisely, let us consider process $x_t(c)$ defined by:

\begin{equation}
x_{t}(c) = \phi(c) x_{t-1}(c) + \eta_t(c) \sqrt{\omega(c) + \alpha(c) x_{t-1}(c)^2},
\label{DARsm}
\end{equation}

\nin We assume that the following condition holds:

\medskip

\nin \textbf{Assumption a.2:}

\nin Let the $L_q$ norm for $q>0$ be denoted by $||.||_q$. Then,

i) For each $c \in (0,1)$, process $\{x_t(c)\}$ is stationary and ergodic.

ii) $c \rightarrow x_t(c)$ is continuous for any $t$ and $||sup_c x_t(c)||_q < \infty$.

iii) There exists $\alpha, 1 \geq \alpha >0$ and
$C_B >0$,  such that 

$|| x_t(c) - x_t(c')||_q < C_B |c-c'|^{\alpha}$ uniformly in $t$ and $c, c' \in (0,1)$.

\medskip

Under Assumptions a.1 and a.2, if $T$ is large and $t/T$ in a small interval $(c-\epsilon, c)$, then the parameters are almost constant over that interval and locally model (\ref{DARsm}) is close to model (\ref{DARloc}) with 
$\phi = \phi(c), \omega=\omega(c), \alpha = \alpha(c)$. This explains the local stationarity.

When all the observations $x_{t,T}, t=1,...,T$ are considered, the variation of the parameters prevents the DAR process from being globally stationary. However, it is locally stationary, if Assumption A.2 is locally satisfied,
i.e. $E [ln | \phi(c) + \eta_t(c) \alpha(c)|] <0$ for any $c$. This is the condition on the negativity of the local Lyapunov exponent.

\subsection{Estimation}

\nin 4.3.1 {\bf Estimation of the Model with Constant Parameters}

The parameter estimates of model (\ref{DAR10}) are obtained by maximizing the quasi-maximum likelihood (QML) objective function, i.e. the log-likelihood function for normally distributed $\eta_t$. 
\begin{equation}
L_T(\theta)= -\frac{1}{2}\sum_{t=2}^T\ln\left(\omega+\alpha x_{t-1}^2\right)-\frac{1}{2}\sum_{t=2}^T\frac{\left(x_t-\phi x_{t-1}\right)^2}{\left(\omega+\alpha x_{t-1}^2\right)},
\label{loglike}
\end{equation}
where $\theta=[\phi,  \  \omega,  \  \alpha]^\prime$. The QML estimators of Model (\ref{DAR10}):
$$ \hat{\theta}_T = Arg max_{\theta \in \Theta} L_T(\theta)$$

\nin are consistent under Assumptions A.1 to A.4,  and the vector of QMLE estimators $\hat\theta_T=[\hat{\phi}_T  \  \hat{\omega}_T  \  \hat{\alpha}_T]^\prime \rightarrow \theta_0$
in probability, where $\theta_0 = [\phi_0  \  \omega_0  \ \alpha_0]^\prime $ [Ling (2004,2007)]. 
Moreover, if the following assumption:

\nin \textbf{Assumption A.5:} $E(\eta_t^4) < \infty$

\nin is satisfied, Li and Ling (2008) and Chen, Li and Ling (2014) show that the Quasi Maximum Likelihood estimators (QMLE) of $\theta$ are also asymptotically normal when $|\phi| \geq 1$ \footnote{The ML/OLS estimators of $\phi$ from a linear autoregressive AR(1) model with constant parameters are not asymptotically normal when $\phi=1$.}
as well as $E(x_t^2) = \infty$. 

$$\sqrt{T} (\hat\theta_T - \theta_0) \rightarrow N(0, diag (\Sigma^{-1}, \kappa \Omega^{-1}))$$

\nin where this convergence is in distribution,  $\Sigma = E_0 [x_{t-1}^2 /(\omega_0 + \alpha_0 x_{t-1}^2)]$  

$$\Omega = E_0 \left( \frac{1}{(\omega_0 + \alpha_0 x_{t-1}^2)^2}\left[ \begin{array}{cc} 1 & x_{t-1}^2 \\ x_{t-1}^2 & x_{t-1}^4 \end{array} \right] \right), $$

\nin $diag (\Sigma^{-1}, \kappa \Omega^{-1})$ denotes the block-diagonal matrix with $\Sigma^{-1}$ as the upper left block and $\kappa \Omega^{-1}$ as the bottom right block and $\kappa$ is the kurtosis less 1 of the distribution of $\eta$. In particular, $\kappa=2$ when $\eta_t$ is normal. These asymptotic results are valid for any true distribution of $\eta$,  not necessarily a Gaussian distribution.
The consistent estimators of $\Sigma$ and $\Omega$  are

$$\hat{\Sigma}_T = \frac{1}{T-1} \sum_{t=2}^T [x_{t-1}^2 /(\hat{\omega}_T + \hat{\alpha}_T x_{t-1}^2)],$$ 

$$\hat{\Omega}_T = \frac{1}{T-1} \sum_{t=2}^T  \frac{1}{(\hat{\omega}_T + \hat{\alpha}_T x_{t-1}^2)^2}\left[ \begin{array}{cc} 1 & x_{t-1}^2 \\ x_{t-1}^2 & x_{t-1}^4 \end{array} \right]. $$

\nin The model residuals  are defined as:

 $$\hat\eta_{t,T}=(x_t-\hat\phi_T x_{t-1})/\sqrt{\hat\omega_T + \hat\alpha_T x_{t-1}^2}.$$ 

\nin The model residuals $\hat{\eta}_{t,T}, t=1,...,T$ allow us to estimate non-parametrically the error density to verify ex-post the symmetry assumption. The parameter $\kappa$ is estimated by $\hat{\kappa}_T = \frac{1}{T-1} \sum_{t=2}^T \hat{\eta}_{t,T}^2 - 1$
$=\frac{1}{T-1}\sum_{t=2}^T\frac{\left(x_t-\hat\phi x_{t-1}\right)^4}{\left(\hat\omega+\hat\alpha x_{t-1}^2\right)^2}-1$
allowing us to accommodate the heavy tailed distribution of the stablecoin prices. 

\bigskip

\nin 4.3.2 {\bf Estimation of the Model with Time-Varying Parameters}

Let us consider the locally stationary tvDAR model. The dynamic model (\ref{DARloc}) of triangular arrays $x_{t,T}, t=1,...,T$ is non-parametric and depends on the functional parameters $\phi(c), \omega(c)>0, \alpha(c)>0, c \in [0,1]$
and on the density function of the noise $\eta_{t,T}$. The estimation of $\phi(.), \omega(.), \alpha(.)$ can be done by the local-in-time QML estimators. We consider a kernel $K$ defined on $ [-1/2,1/2]$ and bandwith $b_T,\,  b_T>0$ (following the notation used in Dahlhaus, Richter, Wu (2019), p. 1035). The local negative log-conditional quasi likelihood is 

\begin{equation}
L_{T,b}(c, \phi, \omega, \alpha)= \frac{1}{Tb_T} \sum_{t=2}^T K \left( \frac{t/T - c}{b_T}  \right) \left[
- \frac{1}{2} \ln \left( \omega+\alpha x_{t-1,T}^2 \right)- \frac{1}{2} \frac{\left(x_{t,T}-\phi x_{t-1,T}\right)^2}{\left(\omega+\alpha x_{t-1,T}^2\right) } \right],
\label{logloc}
\end{equation}

\nin Then, the local negative QML estimator of $\theta(c) = [ \phi(c), \omega(c), \alpha(c)]$ is:

\begin{equation}
\hat{\theta}_{T,b}(c) = Argmax_{\theta} L_{T,b} (c, \phi, \omega, \alpha).
\label{Qloc}
\end{equation}

Under suitable regularity conditions given in [Dahlhaus, Richter, Wu (2019)], this functional QML estimator $\hat{\theta}_{T,b_T}(c), \; c\in [0,1]$ is consistent of $\theta(c), c\in [0,1]$ and its limiting distribution is normal:

$$\sqrt{Tb_T} (\hat{\theta}_{b}(c) - \theta_{0}(c)) \rightarrow N(0, \int K(y)^2 dy \, J(c)^{-1} I(c) J(c)^{-1}),$$

\nin where $\rightarrow$ denotes the weak convergence of processes indexed by $c$, $J(c)$ is the Hessian matrix and $I(c)$ is the outer product of scores, both evaluated at $c$. Under the symmetry assumption on $\eta_t$ and for 
a strictly stationary and ergodic $x_t$, the information matrix $I(c)$ simplifies to a block diagonal matrix [Ling  (2004), Remark 1].

The regularity conditions concern the functional parameter, the distribution of noise $\eta_t$,  the kernel $K$ and the bandwidth $b_T$. They can be found in Dahlhaus, Richter, Wu (2019), since the DAR model is a nonlinear 
autoregressive model discussed in Dahlhaus, Richter, Wu (2019), Example 5.5, p. 1039. In particular, the bandwidth has to satisfy the conditions $ b_T \rightarrow 0, T  b_T \rightarrow \infty, Tb_T^3 \rightarrow 0$ when $T$ tends to infinity.

\subsection{Stability measures}

The Lyapunov exponent measures the average logarithmic rate of separation or convergence of initially close trajectories in chaotic systems, and the sensitivity to initial conditions, in general. 
A negative value of the Lyapunov exponent indicates the stability of the dynamical system, while a positive value indicates chaos. The more negative the Lyapunov exponent, the more stable the system. Therefore, it is used for testing for chaos [Sprott (2003)]. In this section, the Lyapunov exponent is  proposed as a measure of stability of Tether and other stable coins.
In the framework of the DAR model, the Lyapunov exponent is:
$$ \lambda = E(\ln | \phi + \eta \sqrt{\alpha}|).$$

\nin The more negative the Lyapunov exponent, the less explosive the process\footnote{A strictly stationary process can have infinite moments.} and more likely its marginal variance is finite.

The behavior of $E(\ln|\phi + \eta_t \sqrt{\alpha}|)$ as a function of $\phi, \alpha$  can be examined analytically for selected densities of $\eta$, and/or simulated and illustrated graphically [see, Liu et al. (2018) for graphical illustration]. Appendix A.1 shows the analytical formula of $\lambda$ for a uniformly distributed sequence $\{\eta_t\}$. More precisely:

{\bf Proposition 1:} If $\eta \sim U_{[-1,1]}$, the Lyapunov exponent is given by:
$$\lambda(\phi,\alpha) = \frac{1}{2 \sqrt{\alpha}} \left[(|\phi| + \sqrt{\alpha}) \ln( |\phi| + \sqrt{\alpha})
- (|\phi| + \sqrt{\alpha}) - | |\phi| - \sqrt{\alpha}| \ln ||\phi| - \sqrt{\alpha}| + ||\phi| - \sqrt{\alpha}|     \right]$$
\medskip
\nin Proof: See Appendix A.1.

\nin This example clarifies that the Lyapunov exponent is  a continuous function of $\phi, \alpha$, although with points of non-differentiability. Moreover,
it is easy to show that that $E(\ln|\phi + \eta_t \sqrt{\alpha}|)$ is always an even function of $\phi$, i.e. it takes the same value for $\phi$ and $-\phi$. To see that, consider a symmetric density function $\psi(\eta)$. Then,

$$E(\ln|-\phi + \eta_t \sqrt{\alpha}|) = \int (\ln|-\phi + \eta_t \sqrt{\alpha}|) \psi(\eta) d \eta.$$

\nin Because $\psi(\eta)$ is symmetric, we can change the variable $\eta \rightarrow -\eta$:
$$E(ln|-\phi + \eta_t \sqrt{\alpha}|) = \int (\ln|-\phi - \eta_t \sqrt{\alpha}|) \psi(-\eta) d \eta = 
\int (\ln|\phi + \eta_t \sqrt{\alpha}|) \psi(\eta) d \eta = E( \ln|\phi+ \eta_t \sqrt{\alpha}|).$$

\nin This proves that the Lyapunov exponent is even in $\phi$. The Lyapunov exponent can be computed by plug-in from the 
parameter estimates. Let us first consider the constant parameter DAR model.
The following estimators can be considered:

a) Suppose that the true density function $\psi=\psi_0$ is known.  Then, the estimator $ \hat{\lambda}_{1,T}$ of the Lyapunov exponent is:

$$ \hat{\lambda}_{1,T} = \int \ln |\hat{\phi}_T  +\eta \sqrt{\hat{\alpha}_T}| \, \psi_0(\eta)\, d \eta. $$
 
{\bf Proposition 2:}
 
 When the density $\psi(\eta) = \psi_0(\eta)$ is known, then under assumptions A.1 to A.4 and the following condition: 
 
(A.6) $ \;\;\; \exists \delta >0, \mbox{such that} \int sup_{\begin{array}{c} \phi_0 -\delta <\phi < \phi_0 + \delta \\
\alpha_0 -\delta <\alpha < \alpha_0 + \delta \end{array} } |\ln |\phi + \eta \sqrt{\alpha}||
\psi_0 (\eta) d \eta < \infty \;\;\;$

\nin the estimator $ \hat{\lambda}_{1,T}$  converges in probability to the true value  $\lambda_0 = \int \ln|\phi_0 + \eta \sqrt{\alpha}_0| \psi_0(\eta) d\eta$ of the Lyapunov exponent when $T \rightarrow \infty$.

\medskip
\nin Proof: See Appendix A.2

\bigskip

b) When the density $\psi(\eta)$  is unknown, the model is semi-parametric. The Lyapunov exponent can be estimated by the estimator $\hat{\lambda}_{2,T}$ such that:

$$ \hat{\lambda}_{2,T} = \frac{1}{T} \sum_{t=1}^T \ln |\hat{\phi}_T  + \hat{\eta}_{t,T} \sqrt{\hat{\alpha}_T}|. $$

\nin Therefore this estimator is equal to :

$$ \hat{\lambda}_{2,T} = \frac{1}{T} \sum_{t=1}^T \ln \left |\hat{\phi}_T  + \frac{x_t - \hat{\phi}_T x_{t-1}}{\sqrt{\hat{\omega}_T + \hat{\alpha}_T x_{t-1}^2}} \sqrt{\hat{\alpha}_T}\right | = \frac{1}{T} \sum_{t=1}^T g(x_t, x_{t-1}; \hat{\theta}_T)$$

\nin where $g(x_t, x_{t-1}; \theta) = ln \left |\phi  + \frac{x_t - \phi x_{t-1}}{\sqrt{\omega + \alpha x_{t-1}^2}} \sqrt{{\alpha}}\right |$. We also introduce the notation $G_T(\theta) = \frac{1}{T} \sum_{t=1}^T g(x_t, x_{t-1}; \theta)$. 

\medskip
{\bf Proposition 3:} 

Let us introduce the additional conditions:

(A.7)  $ E_{\theta_0} g(x_t, x_{t-1}; \theta) < \infty, \; \forall \theta \in \Theta$.

(A.8) Sufficient Lipschitz condition for stochastic equicontinuity:
There exists a stochastic sequence $B_T$ with $B_T = O_p(1)$  and an increasing function $h$ from $[0, \infty)$ to $[0, \infty)$, continuous at 0 with $h(0)=0$, such that for all $\tilde{\theta}$, $\theta \in \Theta$, $|G_T( \tilde{\theta}) - G_T(\theta)| \leq B_T h(d(\tilde{\theta}, \theta))$.
 
\medskip
\nin Then, under assumptions A.1-A.4 and conditions A.7, A.8 the estimator $ \hat{\lambda}_{2,T} = G_T(\hat{\theta}_T) \rightarrow G(\theta_0) = \lambda_{0} $ in probability.

\medskip
\nin Proof: See Appendix A.3.

\medskip

The asymptotic distributions of estimators $ \hat{\lambda}_{1,T}$, $ \hat{\lambda}_{2,T}$ cannot be obtained asymptotically from the Taylor series expansion because function $\phi, \alpha \rightarrow \int \ln |\phi+\eta \sqrt{\alpha}| \psi(\eta) d\eta$ does not satisfy the necessary differentiability assumption, as pointed out in Proposition 1. However, the distribution  of $ \hat{\lambda}_{1,T}$, $ \hat{\lambda}_{2,T}$ can be determined by simulations and used for hypothesis testing.

\medskip

c) An alternative stability measure is  $\xi =\phi^2 + \alpha$, which depicts the region of parameter space $\xi < 1$ where the marginal variance of $x_t$ remains finite, so that the process is both strictly and weakly stationary. In that sense $\xi$ is a more conservative measure of stability than the Lyapunov exponent because  there is a region of parameter values $\phi, \alpha$  where the condition $\xi <1$ no longer holds, while the condition $\lambda <0$ remains satisfied. The estimator $\hat{\xi}_T$ of $\xi$ is:

$$\hat{\xi}_T = \hat{\phi}_T^2 + \hat{\alpha}_T.$$

\medskip

{\bf Proposition 4:} 

Under Assumptions A.1-A.5, the estimator $\hat{\xi}_T$ converges in probability to $\xi_0$ when  $T \rightarrow \infty$ and it is asymptotically Normally distributed:

$$ \sqrt{T} (\hat{\xi}_T - \xi_0) \stackrel{\rm A}{\rm \sim} N(0, V_{\xi}),$$ 

\nin where the formula of the asymptotic variance $V_{\xi}$ is given in Appendix A.4. 

The asymptotic variance  provides the asymptotically valid standard errors that can be used to test the null hypothesis $\xi< \xi_0$ using a Wald test statistic. The interpretation of measure $\xi$ is similar to that of $\lambda$: the smaller $\xi$, the more stable $x_t$.

\nin {\bf 4.4.1 Model with time varying parameters}

The Lyapunov exponent $\lambda_2(c)$ can be estimated locally by computing $\hat{\lambda}_{2,T}(c)$ from the plugged in parameter estimates and residual values of the tvDAR model. For example, 
the Lyapunov exponent can be estimated from a kernel-weighted formula 

$$\hat{\lambda}_{2,T}(c) =  \frac{1}{Tb_T} \sum_{t=1}^T K \left( \frac{t/T - c}{b_T}  \right) (ln| \hat{\phi}(t/T) + \hat\eta_{t,T} \sqrt{\hat\alpha(t/T)}|), $$

\nin  with 
the Epanechnikov kernel $K(c)=\frac{3}{2}(1-(2c)^2)$ for $c \in [-1/2,1/2]$ and $K(c)=0$ otherwise, which satisfies the regularity condition for localizing kernel [see Dahlhaus, Richter, Wu (2019, Assumption 2.6)].

A similar approach can be used to estimate locally the measure $\xi(c) = \phi^2(c) +\alpha(c)$. The local plug-in estimator $\hat{\lambda}_{2,T}(c)$ is illustrated in the next section.

\section{Empirical Results}

This section presents the parameter estimates for the constant parameter DAR model and the time varying parametr tvDAR model.
The time varying parameters DAR model is estimated first by rolling, which is equivalent to the use
of an asymmetric rectangular kernel, as shown in the previous section. 
Next, the model is estimated by using a kernel which assigns higher weights to the observations close to the estimation date, providing consistent and asymptotically normally distributed estimates, which are used for testing hypotheses on the constancy of parameters. The estimators of stability measures are also computed and illustrated graphically.

\medskip

The estimation of the DAR(1) process with constant parameters is straightforward. First, we demean the price series of Tether by subtracting the total mean of 1.0022 and then fit the model \ref{DAR10} to the entire series of the demeaned prices, which gives the following result

\bigskip

\begin{table}[htpp]
  \centering
\scalebox{0.85}{  
\begin{threeparttable}
  \caption{Estimation of the DAR(1) model using the entire sample \label{tab:constantmodel}}
    \begin{tabular}{rccccc}
    \toprule
    \toprule
	& \multicolumn{5}{c}{DAR(1) parameters} \\
          & $\phi$   &       & $\omega$ &       & $\alpha$ \\
\cmidrule{2-2}\cmidrule{4-4}\cmidrule{6-6}          &       &       &       &       &  \\
    \multicolumn{1}{l}{Estimates} & 0.699 &       & $9.102e^{-06}$ &       & 0.484 \\
          &       &       &       &       &  \\
    \multicolumn{1}{l}{Standard deviation} & (0.034) &       & ($2.174e^{-06}$) &       & (0.205) \\
          &       &       &       &       &  \\
    \bottomrule
    \end{tabular}%
 \end{threeparttable}}
\end{table}%

\nin where the standard deviations of the parameters are obtained using the asymptotic distribution described in Section 5.3.1.

In the following sections, the model \ref{DARloc} is estimated by using two types of kernels. The first one is the asymmetric rectangular kernel, equivalent to the rolling estimation over the window of 50. This window ensures good properties of the estimators, while preventing the over-smoothing.

The second approach relies on the Epanechnikov kernel and produces consistent and asymptotically normally distributed parameter estimates  used for hypotheses testing. 

\subsection{Rectangular kernel}

The tvDAR(1) is estimated from the demeaned Tether price by rolling, which is a common practice in applied literature, and a window of length 50 days. This is equivalent to using an asymmetric rectangular kernel, $K(u) = 1_{(-1,0)}(u), \; b_T = 50/T$ which is computationally simple, but does not satisfy the smoothness conditions ensuring locally the validity of asymptotic distribution of the QMLE.  The rolling estimate and the confidence interval of the parameter of interest $\phi(t/T)$  is displayed in the first paned of Figure \ref{DAR1} below.

 \begin{figure}[H]
\centering
  \centering
  \includegraphics[width=15cm,height=12cm,angle = 0]{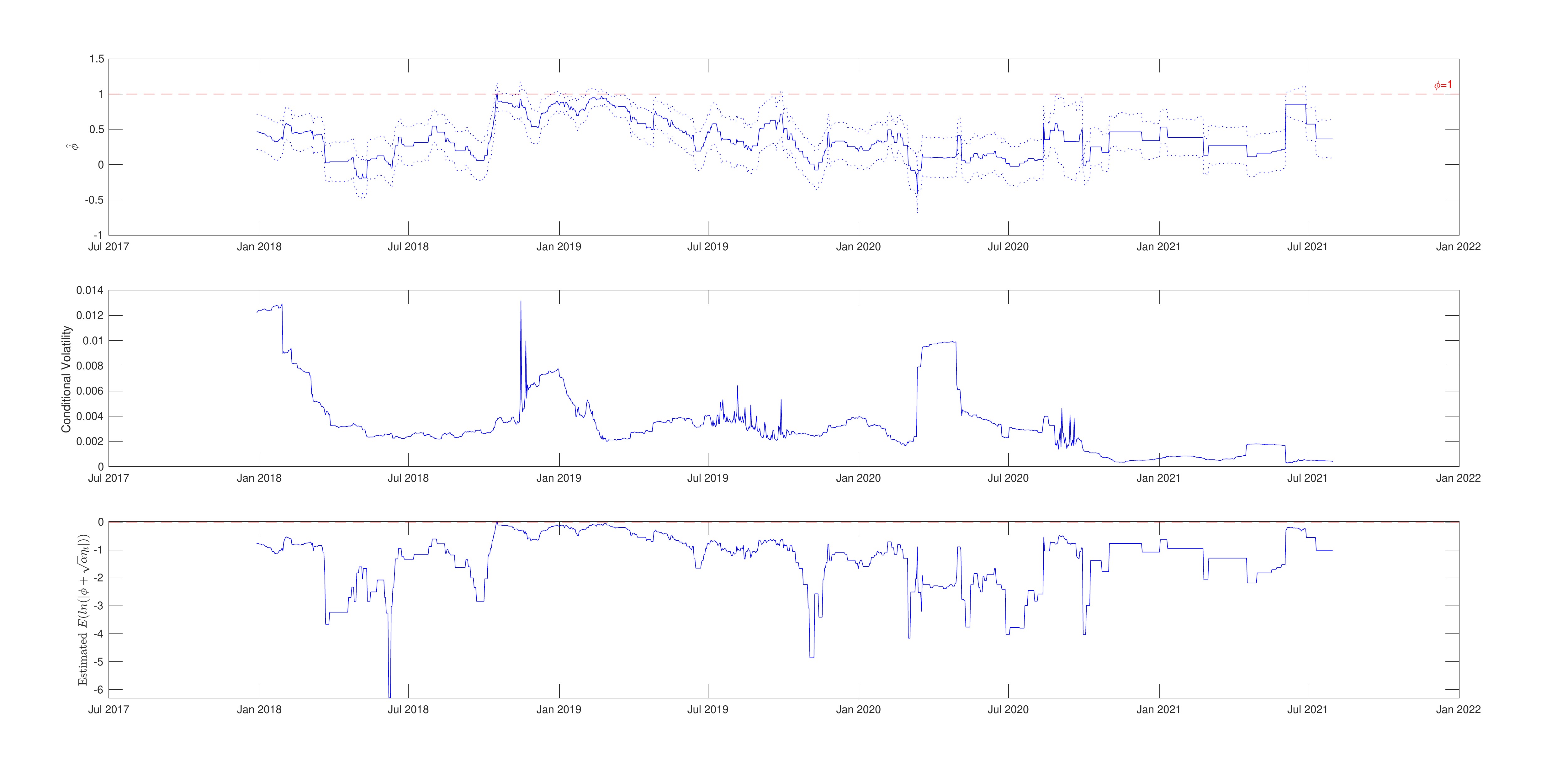}
\caption{tvDAR(1) parameter $\phi(t/T)$, conditional volatility and Lyapunov exponent $\lambda_2(t/T)$}
 \label{DAR1}
  \end{figure}

The second panel shows the estimated conditional volatility  $\sqrt{\hat\omega(t/T) + \hat\alpha(t/T) x_t^2}$ for Tether price.  The third panel in  Figure \ref{DAR1} presents the local estimates $\hat{\lambda}_{2,T}(t/T)$ of the Lyapunov exponent
computed by plugging in the local parameter estimators.

We observe that there are periods when the autoregressive coefficient is not significantly different from 1. For instance, this is the case between October and December 2018 and between February and May 2019. These results from the tvDAR (1) model confirm our initial observation in Section 3.1 that the price series of Tether shows strong persistence in 2018 and 2019 whilst allowing us to pinpoint its exact timing. The estimated conditional volatility based on the estimated parameters has a pattern consistent with the local variance estimator in Figure \ref{TMV}. The results in the third panel suggests that Assumption 2 holds and the Lyapunov exponent remains negative even for the highest recorded persistence. The sample Lyapunov exponent varies across time becoming on average more negative before the end of the sampling period. This indicates that Tether achieves higher stability over that period.

 \begin{figure}[H]
\centering
    \begin{subfigure}[t]{\textwidth}
  \centering
     {\caption*{Full Sample}}
  \includegraphics[width=15cm,height=4.5cm,angle = 0]{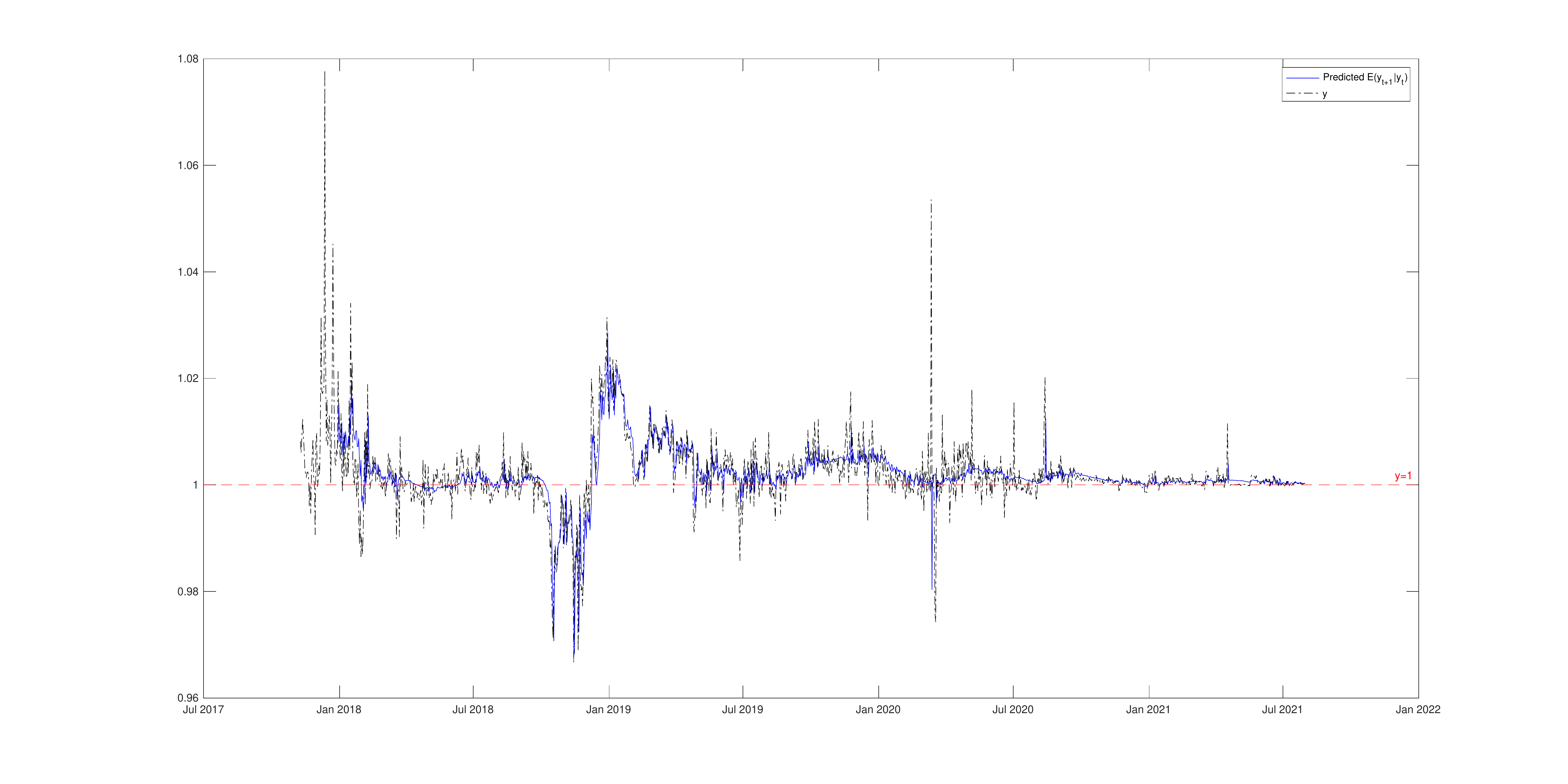}
  \end{subfigure}
  
    \begin{subfigure}[t]{\textwidth}
  \centering
      {\caption*{October 2018 to October 2019}}
\includegraphics[width=15cm,height=4.5cm,angle = 0]{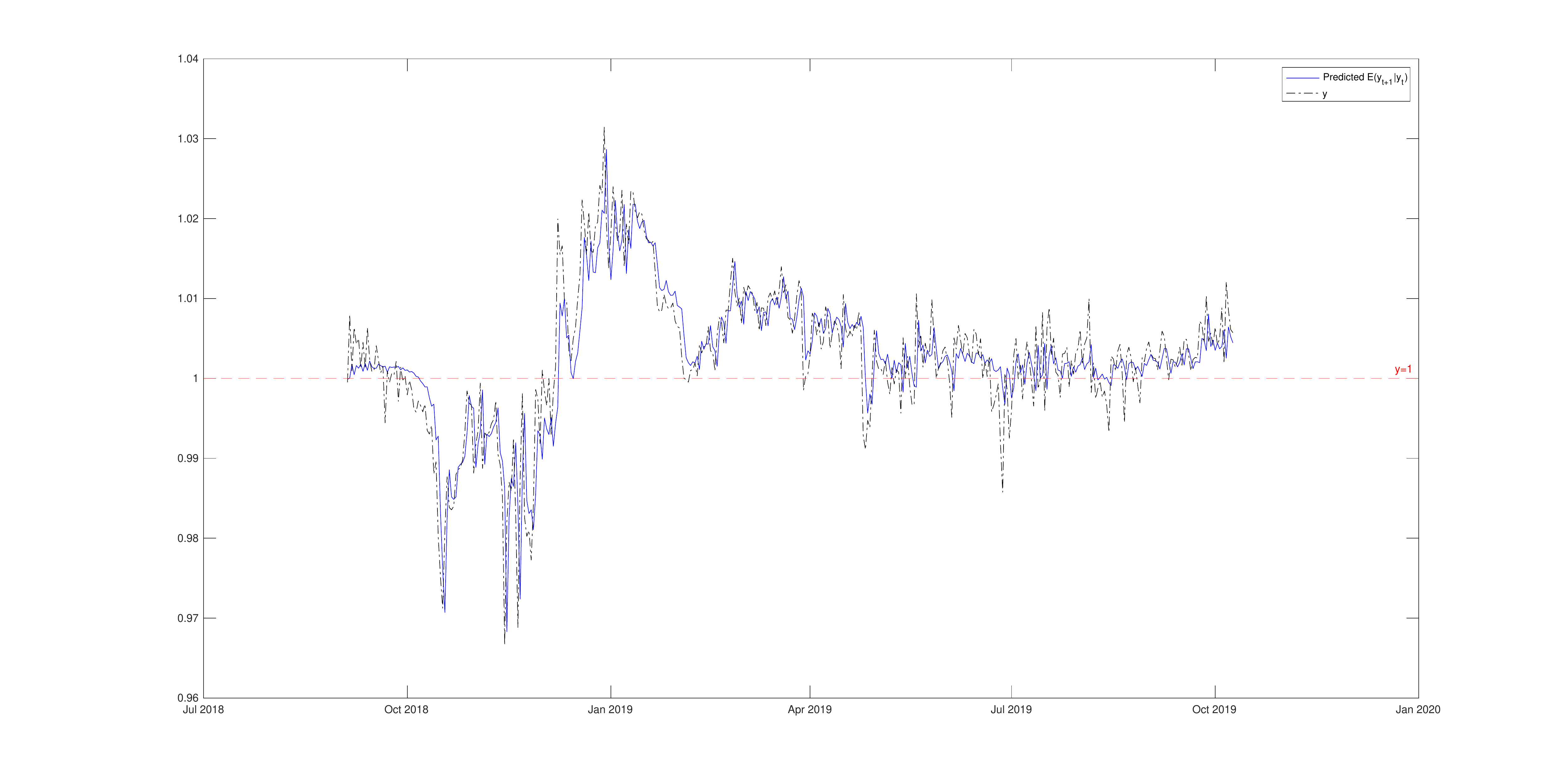}
  \end{subfigure}
  
\caption{Tether prices compared to one-step ahead out-of-sample forecasts}
 \label{FDAR0}
  \end{figure}
  
\begin{figure}[H]
\centering
    \begin{subfigure}[t]{\textwidth}
  \centering
     {\caption*{Full Sample}}
  \includegraphics[width=15cm,height=5cm,angle = 0]{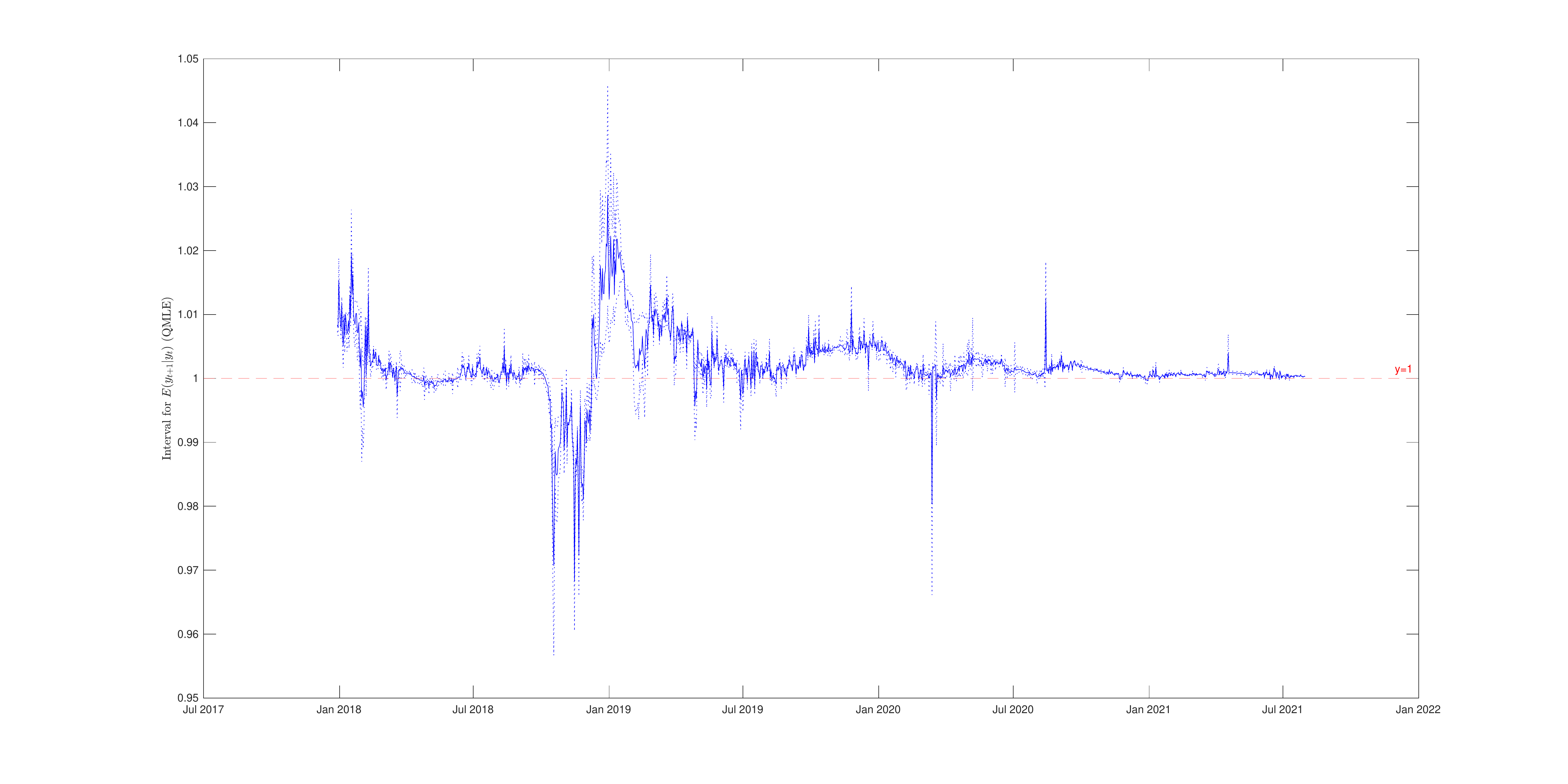}
  \end{subfigure}
  
    \begin{subfigure}[t]{\textwidth}
  \centering
      {\caption*{October 2018 to October 2019}}
\includegraphics[width=15cm,height=5cm,angle = 0]{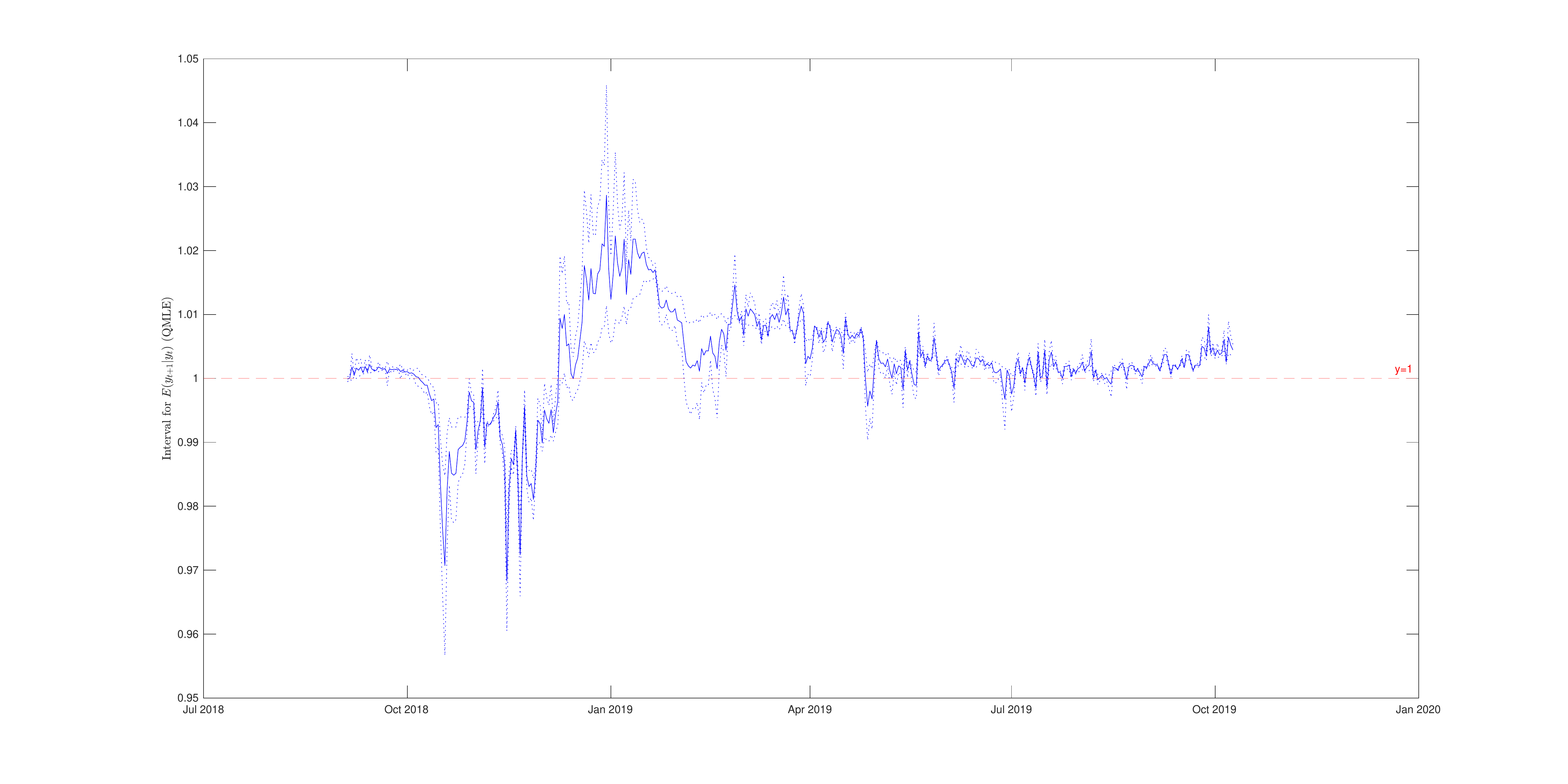}
  \end{subfigure}
  
\caption{One-step ahead out-of-sample predicted Tether prices and prediction  intervals} 
 \label{FDAR}
  \end{figure}

 The tvDAR(1) model estimated with the asymmetric rectangular kernel can be used for forecasting at short horizons, under the assumption that the parameter functions
remain constant and equal to the last estimated value.
Figure \ref{FDAR0} presents the observed Tether price and  the estimate $\hat y_{t+1}$ of the one-day-ahead conditional mean $E(y_{t+1}|y_t,\ldots)$ of the price of Tether using a rolling window. To get $\hat y_{t+1}$, we add the local mean of Tether price to the estimated $\hat \phi$ using data over $50$ days up to date $t$ times $x_t$. The figure shows a close match between Tether price and its best prediction based on the previous day price. In addition, the computed mean square prediction error is $1.7428 \times 10^{-5}$ which is very small. Under the assumption of locally constant parameters, the 95 \% asymptotically valid prediction intervals in Figure \ref{FDAR} are given by $\left[\hat y_{t+1}- \frac{1.96}{\sqrt{T}}\sqrt{x_t^2\hat{\Sigma}^{-1}}, \  \hat y_{t+1}+ \frac{1.96}{\sqrt{T}} \sqrt{x_t^2\hat{\Sigma}^{-1}}\right]$.

Next, we conduct further investigations to analyze the goodness of fit of the tvDAR(1). For this analysis, we keep the estimation window of $50$ and perform the Ljung-Box test of white noise on $\hat\eta_t$ and $\hat\eta_t^2$, while rolling the sample used for the test [see Li (1992), and Li and Mak (1994)].   Because we consider all subsamples of $50$ consecutive dates, it is likely that at some dates the serial correlation  is not fully captured by the tvDAR(1) model. Our results show that for most periods, residuals $\hat\eta_t$ and $\hat\eta_t^2$  are serially uncorrelated. More precisely, we reject the null of no serial correlation for $\hat\eta_t$ only for $1.52\%$ of the subsamples, while we reject the null of no serial correlation for $\hat\eta_t^2$ only for $8.08\%$ of the subsamples. The results suggest that the model captures most of the nonlinear serial correlation in Tether prices. 

\subsection{Epanechnikov kernel}  

We now use a symmetric Epanechnikov kernel producing consistent and asymptotically normally distributed estimates
for hypothesis testing.

\begin{figure}[H]
\centering
  \centering
  \includegraphics[width=15cm,height=10cm,angle = 0]{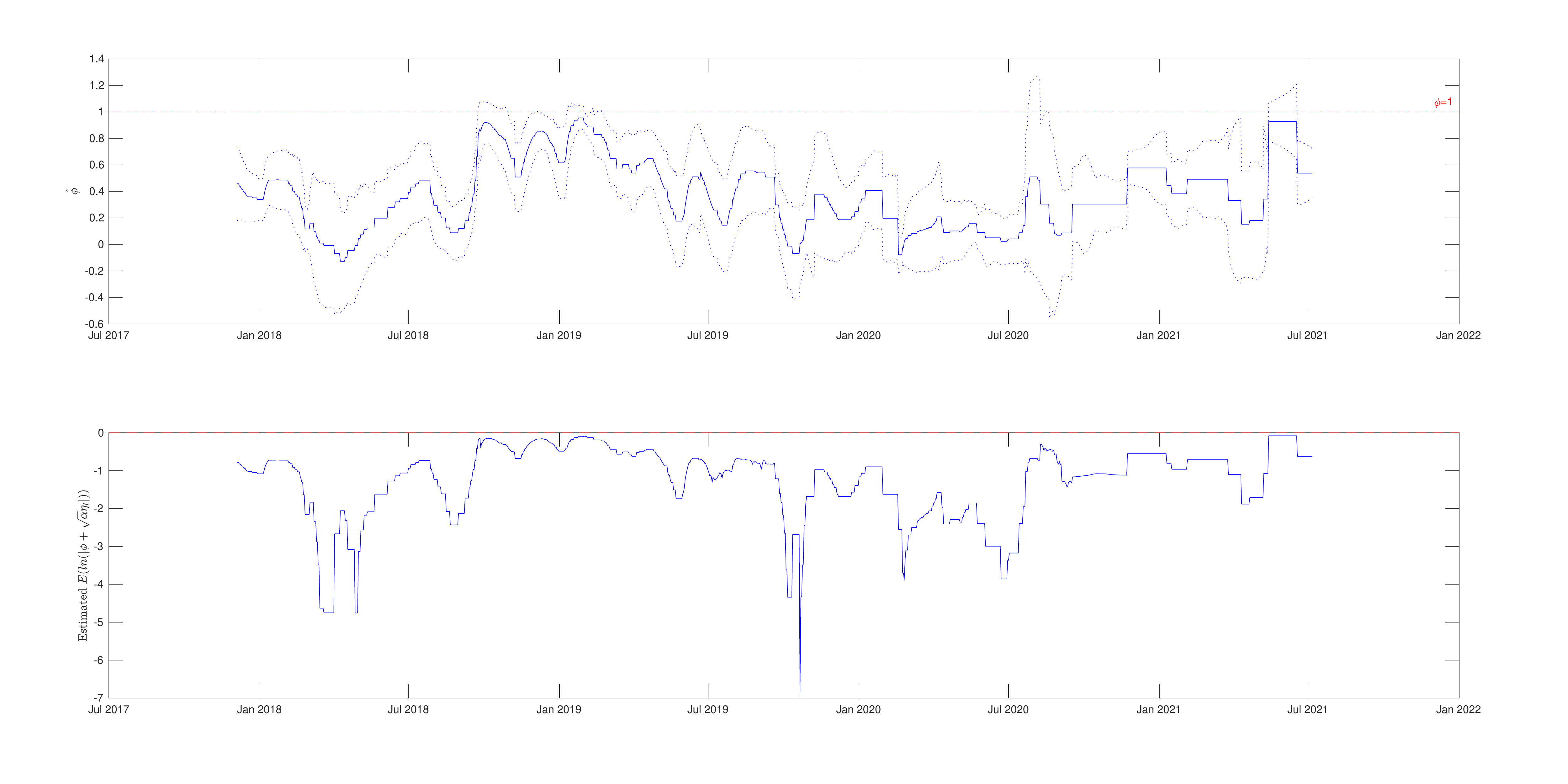}
\caption{Kernel-based parameter $\phi$ estimates and Lyapunov exponent $\lambda(t/T)$}
 \label{DAR2}
 \end{figure}

The first panel of Figure \ref{DAR2} plots the estimated time-varying autoregressive coefficient and its confidence band, while the second panel  presents the time-varying estimates for of the Lyapunov exponent using the $\hat{\lambda}_{2,T}(t/T)$ estimator. For the bandwidth, we choose $b_T=50/T$ using the same window as before.

The results confirm that when the estimation is conducted locally over each period, the Lyapunov exponent displayed in the second panel of Figure \ref{DAR2} is negative. Hence the critical validity condition holds for all the dates. Moreover, the Lyapunov exponent is on average more negative at the end of the sampling period, confirming that Tether has achieved higher stability at the end  of the sampling period. 

Furthermore, the estimated dynamic of estimated parameter $\phi$ for Tether price in the first panel of of Figure \ref{DAR2} remains similar to that of Figure \ref{DAR1}. These findings confirm that Tether price has causal dynamics, as the autoregressive parameter and its confidence interval are mostly between $-1$ and $1$. However, there are few dates when this estimate is not statistically different from $1$, which suggests strong persistence in Tether prices.

To detect the periods of strong persistence, we test the null hypothesis $H_0 : \phi = 1 $  using the series  of $\hat\phi(t/T)$ and its 95\% confidence intervals for each $t=1,...,T$. We conduct the hypothesis test using the series of  $\hat\phi(t/T)$ obtained from the estimation with the Epanechnikov kernel and report the results in Table \ref{tab:episodes}.

The results suggest that although Tether price is predictable most of the time, there are intervals of time periods where this tends not to be the case. During these episodes presented in Table \ref{tab:episodes}, we find high persistence in Tether price. However, as explained above, the proposed tvDAR approach remains valid and accommodates strong persistence in the price series. In light of the results in Figure \ref{events}, we further inspect the identified periods. We observed that the identified episodes in 2018 and 2019 overlap with the periods of high volatility observed in Figure \ref{events}, which ended in February 2020, but started around the introduction in September 2018 of  USD Coin, another stablecoin designed to maintain price equivalence to the U.S. dollar. Moreover, the episodes in 2020 match with a small rise in Tether price volatility by the end of July 2020, while the episodes in 2021 can be associated with the period of increased volatility at the end of our sample.

\begin{table}[H]
  \centering
    \caption{Episodes of high persistence in Tether characterized by $\phi=1$.    \label{tab:episodes}}
    \begin{tabular}{ccccc}
    \toprule
    \textit{Year} & & \textit{From} &  &\textit{To}\\
               &       &  &       & \\
       \midrule
                  &       &  \\
    2018 & & September 24 & & October 29 \\
            &  & December 3 &  & October 5 \\
              &       &        &       &  \\
     \midrule
    &       &           &       &  \\
    2019 &  & January 11 &  & January 30\\
          &   & February 9 &  & February 11\\
          &    & February 16 &  & February 21\\    
                         &       &  &       &  \\     
        \midrule 
               &       &    &       &  \\
    2020 & & July 25 &  & August 8\\
            &  & August 16 & & August 19\\
                &       &        &       &  \\
         \midrule
             &       &       &       &  \\
    2021 & & May 14 & & June 17\\  
               &       &   &       & \\       
      \bottomrule
    \end{tabular}%
\end{table}%

\subsection{Test for Conditional Homoscedasticity}  

Let us now consider a simple test of model specification of the tvDAR(1) model with time-varying parameters.
It is based on testing for the constancy of the variance function in model \ref{DARloc} which is given by the following expression

\begin{equation}
\sigma(t/T) =  \sqrt{\omega(t/T) + \alpha(t/T) x_{t-1,T}^2} , \quad  t=1,...,T.
\label{variancefunc}
\end{equation}

\medskip

To test the null hypothesis $H_0:\sigma(t/T)=\sigma_0 \: \forall t $, we consider the below test statistics proposed by Chandler and Polonik (2017) for time-varying autoregressive processes

\begin{equation}
CP_T = \sup_{\alpha \in [0,1]} \sqrt\frac{T}{(\gamma(1-\gamma)}  \lvert \hat G_{T,\gamma}(\alpha) - \alpha \gamma \rvert,
\label{teststatistics}
\end{equation}

\medskip

where

\begin{itemize}

\item $ \hat G_{T,\gamma}(\alpha) = \frac{1}{T} \sum_{t=1}^{[\alpha T]}\mathbbm{1}(\hat \epsilon^2_t \geq \hat q^2_\gamma),$

\item $ \hat q^2_\gamma = \min \left( q^2 \geq 0 : \frac{1}{T} \sum_{t \in [aT, bT]} \mathbbm{1}(\hat \epsilon^2_t > q^2) \leq \gamma \right),$

\item $\hat \epsilon_t = x_t - \hat \phi ( \frac{t}{T}) x_{t-1} .$

\end{itemize}

The process $ \hat G_{T,\gamma}(\alpha) $ counts the number of squared residuals within the first $(100 \times \alpha) \%$ of the observations that are larger than the empirical quantile of the squared residuals denoted by $\hat q_\gamma^2$. The series of squared residuals is constructed by computing $\hat \epsilon_t = x_t - \hat \phi ( \frac{t}{T}) x_{t-1} $ for each period t where $\hat \phi ( \frac{t}{T})$ in our case is the corresponding DAR(1) estimate obtained in the previous section.

Chandler and Polonik (2017) shows that the test statistics $CP_T$ in equation (\ref{teststatistics}) under the null hypothesis converges asymptotically to the supremum of a Brownian bridge. This asymptotic result is found to be still valid when the time-varying functions of the model parameters are estimated nonparametrically (see Chandler and Polonik (2012)).

When computing the test statistics $CP_T$, we consider different alternatives for the empirical upper $\gamma$-quantile for comparison, particularly $\gamma = (0.7, 0.8, 0.9)$ following Chandler and Polonik (2017). Given the choice of $\gamma$,  we then calculate the expression $\sqrt\frac{T}{(\gamma(1-\gamma)}  \lvert \hat G_{T,\gamma}(\alpha) - \alpha \gamma \rvert$ for different values of $\alpha$  and report the results in table \ref{tab:teststats}.

\begin{table}[htbp]
  \centering
\scalebox{0.85}{  
\begin{threeparttable}
  \caption{The calculated values of $\sqrt\frac{T}{(\gamma(1-\gamma)}  \lvert \hat G_{T,\gamma}(\alpha) - \alpha \gamma \rvert$ for the given pairs of $\gamma$ and $\alpha$.  \label{tab:teststats}}
   
    \begin{tabular}{rrcccccccccc}
    \toprule
          &       & \multicolumn{10}{c}{$\alpha$} \\
          &       & 0.1   & 0.2   & 0.3   & 0.4   & 0.5   & 0.6   & 0.7   & 0.8   & 0.9    \\
          &       &       &       &       &       &       &       &       &       &       &  \\
\cmidrule{3-12}          &       &       &       &       &       &       &       &       &       &        \\
          & \multicolumn{1}{c|}{0.7} & 0.922 & 1.302 & 2.466 & 3.328 & 4.251 & 5.957 & 6.759 & 6.837 & 3.539  \\
          & \multicolumn{1}{c|}{} &       &       &       &       &       &       &       &       &       \\
    \multicolumn{1}{l}{$\gamma$} & \multicolumn{1}{c|}{0.8} & 0.912 & 1.478 & 2.390 & 3.233 & 3.937 & 5.471 & 6.106 & 6.327 & 3.509  \\
          & \multicolumn{1}{c|}{} &       &       &       &       &       &       &       &       &        \\
          & \multicolumn{1}{c|}{0.9} & 0.562 & 1.031 & 1.593 & 2.155 & 2.993 & 3.923 & 4.485 & 5.047 & 3.490  \\
          &       &       &       &       &       &       &       &       &       &        \\
    \bottomrule
    \end{tabular}%
 \end{threeparttable}}
\end{table}%

The table shows that regardless of the choice of $\gamma$, the largest value is achieved when $\alpha = 0.8$. By the definition of supremum, the test statistics $CP_T$ should satisfy $CP_T \geq \sqrt\frac{T}{(\gamma(1-\gamma)}  \lvert \hat G_{T,\gamma}(\alpha) - \alpha \gamma \rvert$ for all $\alpha \in [0,1]$. Hence, it is safe to say that we have $CP_T \geq 5.047$ in the worst case scenario, i.e when $\gamma=0.9$ is chosen.\footnote{Alternatively, we could fix our choice of $\gamma$ and find the exact value of $\alpha \in [0,1]$ at which the expression $\sqrt\frac{T}{(\gamma(1-\gamma)} \lvert \hat G_{T,\gamma}(\alpha) - \alpha \gamma \rvert$ attains its maximum on this fixed interval. This could slightly improve the precision of the lower bound for the test statistics. For example, when $\gamma=0.9$, the expression attains its maximum value of 5.106 at $\alpha^*=0.8012$.} Compared with the critical values from the asymptotic distribution of the test statistics,\footnote{see https://homepages.ecs.vuw.ac.nz/~ray/Brownian/ for the distribution of the supremum of a Brownian Bridge.} this result leads us to the conclusion that we can reject the null hypothesis of constant variance function even at the 99\% confidence level. In other words, we have a statistically significant evidence in favor of the alternative that the variance function is varying over time, which consequently justifies our strategy to estimate the model with time-varying parameters.


\section{Concluding Remarks}
We show that the distributional and the dynamic properties of stablecoins have been evolving over the sampling period. We implement local analysis to detect and describe local explosive
patterns, time-varying volatility and persistence. We model the dynamic of the most important stablecoin which is Tether, and provide evidence that the tvDAR(1) model with time varying coefficients
provides locally a good fit and reliable short-term predictions of Tether prices. Our modelling strategy enables us to have valid inference even when the tvDAR(1) coefficient $\phi_t$ of Tether price is not locally different from 1.
The sample Lyapunov exponent computed from the parameter estimates of the  model provides a measure of stability. It confirms that at the end of the sampling period Tether becomes relatively more stable and allows for comparing the stability of Tether with other stablecoins.

\bigskip

\setcounter{section}{0}\def\thesection{A}
\setcounter{subsection}{0}
\section*{Appendix A:  Technical Results}

This Appendix contains the proofs of Propositions 1, 2, 3 and 4.

\subsection{Proposition 1}

Because of the symmetry of the density of $\eta$, the Lyapunov exponent is an even function of $\phi$. Hence we can suppose that $\phi >0$ to find the expression of the Lyapunov exponent, and then replace $\phi$ by $|\phi|$.

\nin For $\phi>0$ we have:

$$ \lambda(\phi, \alpha) = E \ln |\phi + \eta \sqrt{\alpha}| = \int \ln |\phi + \eta \sqrt{\alpha}| \psi(\alpha) d \alpha.$$

\nin Let us assume that $\eta \sim U_{[-1,1]}$. Then, its density is $\psi(\eta) = \frac{1}{2} 1_{\eta \in [-1,1]}$.

\nin We have:

$ \lambda(\phi, \alpha)  = \frac{1}{2} \int_{-1}^1 \ln |\phi + \eta \sqrt{\alpha}| d \alpha$.

\nin We observe that: 

$$
\begin{array}{ll}
\phi + \eta \sqrt{\alpha} >0 & \iff \eta > - \phi/\sqrt{\alpha}, \\
  \phi + \eta \sqrt{\alpha} <0 & \iff \eta < - \phi/\sqrt{\alpha}.
\end{array}
$$

\nin Hence,
$$ \lambda(\phi, \alpha) = \frac{1}{2} \int_{-1}^1 1_{\eta > - \phi/\sqrt{\alpha}} \ln (\phi + \eta \sqrt{\alpha}) d \eta + \frac{1}{2} \int_{-1}^1 1_{\eta < - \phi/\sqrt{\alpha}} \ln (-\phi - \eta \sqrt{\alpha}) d \eta$$

\nin Let us now examine the two cases:

a) If $\phi/\sqrt{\alpha} >1 \iff - \phi/\sqrt{\alpha} <-1$, we get:

$
\begin{array}{ll}
\lambda(\phi, \alpha) & =\frac{1}{2} \int_{-1}^1  \ln (\phi + \eta \sqrt{\alpha}) d \eta + \frac{1}{2}0 
= \frac{1}{2} \frac{1}{\sqrt{\alpha}} \int_{-1}^1  \ln (\phi + \eta \sqrt{\alpha}) d (\sqrt{\alpha} \eta) \\ 
& = \frac{1}{2\sqrt{\alpha}} \int_{\phi -\sqrt{\alpha}}^{\phi +\sqrt{\alpha}}  \ln (u) d u 
\mbox{ with change of variable} \;\; u = \phi + \eta \sqrt{\alpha}  \\ 
& =  \frac{1}{2\sqrt{\alpha}} u \ln(u) -u |_{\phi -\sqrt{\alpha}}^{\phi +\sqrt{\alpha}} 
\end{array}
$

$ = \frac{1}{2\sqrt{\alpha}}\left[(\phi+\sqrt{\alpha}) \ln (\phi+\sqrt{\alpha}) - (\phi+\sqrt{\alpha}) - (\phi-\sqrt{\alpha}) \ln (\phi-\sqrt{\alpha}) + (\phi-\sqrt{\alpha})\right]
$
\medskip

b) If $\phi/\sqrt{\alpha} <1  \iff -\phi /\sqrt{\alpha} > -1$, we get:

$
\begin{array}{l}
\lambda(\phi, \alpha ) = \frac{1}{2} \int_{-\phi /\sqrt{\alpha}}^1  \ln (\phi + \eta \sqrt{\alpha}) d \eta +  \frac{1}{2} \int_{-1}^{-\phi /\sqrt{\alpha}}  \ln (-\phi - \eta \sqrt{\alpha}) d \eta  \\
= \frac{1}{2} \frac{1}{\sqrt{\alpha}} \int_{0}^{\phi+\sqrt{\alpha}}  \ln (u) du +\frac{1}{2} \int_{\phi /\sqrt{\alpha}}^1  \ln (-\phi + \eta \sqrt{\alpha}) d \eta  \mbox{ with the change of variable} \; u = \phi+\eta \sqrt{\alpha} \\
= \frac{1}{2\sqrt{\alpha}} \int_0^{\phi +\sqrt{\alpha}}  \ln (u) d u +
  \frac{1}{2\sqrt{\alpha}} \int_0^{-\phi +\sqrt{\alpha}} \ln(u) du
\end{array}
$

$ = \frac{1}{2\sqrt{\alpha}}\left[(\phi+\sqrt{\alpha}) \ln (\phi+\sqrt{\alpha}) - (\phi+\sqrt{\alpha}) - [(-\phi+\sqrt{\alpha}) \ln (-\phi+\sqrt{\alpha})] + (-\phi + \sqrt{\alpha})\right]
$

\medskip
\nin By putting the two expressions in a) and b) together we get for $\phi>0$:

$$
\lambda(\phi, \alpha) = \frac{1}{2\sqrt{\alpha}}\left[(\phi+\sqrt{\alpha}) \ln (\phi+\sqrt{\alpha}) - (\phi+\sqrt{\alpha}) - |\phi-\sqrt{\alpha}| \ln |\phi-\sqrt{\alpha})| + |\phi - \sqrt{\alpha}| \right]
$$

\nin The general expression of the Lyapunov Exponent without the sign constraint on $\phi$ is: 

$$
\lambda(\phi, \alpha) = \frac{1}{2\sqrt{\alpha}}\left[ (|\phi|+\sqrt{\alpha}) \ln (|\phi|+\sqrt{\alpha}) - (|\phi|+\sqrt{\alpha}) - ||\phi|-\sqrt{\alpha}| \ln ||\phi|-\sqrt{\alpha})| + ||\phi| - \sqrt{\alpha}| \right]
$$

\nin We observe a non-differentiability in $\phi = \pm \sqrt{\alpha}$. Moreover, for $\phi=0$, we get a zero value of the Lyapunov exponent:

$$\lambda(\phi, \alpha) = \frac{1}{2\sqrt{\alpha}} \left[ \sqrt{\alpha} \ln \sqrt{\alpha}
- \sqrt{\alpha} \ln \sqrt{\alpha} + \sqrt{\alpha} \right] = 0$$

\subsection{Proof of Proposition 2}

We need to show that when $T \rightarrow \infty$, then
$\int \ln [\hat{\phi}_T + \eta \sqrt{\hat{\alpha}_T}] \psi_0(\eta) d \eta \rightarrow 
\int \ln (\phi_0 + \eta \sqrt{\alpha_0}) \psi_0 (\eta) d \eta$ in probability if
$(\hat{\phi}_T, \hat{\alpha}_T) \rightarrow (\phi_0, \alpha_0)$ in probability, and if condition
A.6 of Proposition 2:

$$ \exists \delta >0, \mbox{such that} \int sup_{\begin{array}{c} \phi_0 -\delta <\phi < \phi_0 + \delta \\
\alpha_0 -\delta <\alpha < \alpha_0 + \delta \end{array} } |\ln |\phi + \eta \sqrt{\alpha}||
\psi_0 (\eta) d \eta < \infty \;\;\; (\mbox a.1)$$

\nin holds.

\medskip

 Proof of convergence:

\nin If $(\hat{\phi}_T, \hat{\alpha}_T) \rightarrow (\phi_0, \alpha_0)$ in probability, then they also converge in distribution. It follows from the Skorokhod theorem that up to a change of probability space, we can assume that the almost sure (a.s.) convergence also holds [Billingsley (1999)]. Therefore, if $g$ is a continuous function of $(\phi, \alpha)$, we have $g(\hat{\phi}_T, \hat{\alpha}_T) \rightarrow g(\phi_0, \alpha_0)$ a.s. and in distribution in that new space. Then, $g(\hat{\phi}_T, \hat{\alpha}_T) \stackrel{\rm d}{\rm \rightarrow}g(\phi_0, \alpha_0)$ in the initial space and also in probability because the limit is constant, we get the "in probability" version  of the continuous mapping theorem. Therefore, we need only  the condition ensuring that $g(\phi, \alpha) = \int \ln | \phi + \eta \sqrt{\alpha}| \psi_0 (\eta) d\eta$ is continuous. Condition (A.6) ensures the continuity of integral function $g$, which follows from the dominated convergence theorem.

\subsection{Proof of Proposition 3}

We first prove a general lemma, which is next applied to the DAR model and Lyapunov estimator $\lambda_{2,T}$.

{\bf Lemma}

Let us consider a sequence $G_T(\theta)$ of stochastic functions of $\theta$, $\theta \in \Theta$, and a sequence of estimators $\hat{\theta}_T$. We assume that:

i) $\Theta$ is compact and $\theta_0$ is in the interior of $\Theta$.

ii) $\hat{\theta}_T$ tends in probability to $\theta_0$.

iii) $G_T (\theta)$ tends in probability to a limit $G(\theta), \; \forall \theta \in \Theta$.

iv) Sufficient Lipschitz condition for stochastic equicontinuity: 

There exists a stochastic sequence $B_T$ with $B_T = O_p(1)$ and an increasing function $h:$ $[0, \infty) \rightarrow [0, \infty)$ continuous at zero, with $h(0) = 0$ and such that for all $\tilde{\theta}$, $\theta \in \Theta$, $|G_T( \tilde{\theta}) - G_T(\theta)| \leq B_T h(d(\tilde{\theta}, \theta))$.

\nin Then, $\hat{G}_T (\hat{\theta}_T)$ tends in probability to $G(\theta_0)$. 
 
\medskip
\nin Proof:

\nin We have :

$ \begin{array}{ll}
|G_T (\hat{\theta}_T) - G(\theta_0)| & = |G_T (\hat{\theta}_T) - G_T(\theta_0) + G_T(\theta_0) - G(\theta_0)| \\
& \leq |G_T (\hat{\theta}_T) - G(\theta_0) | + | G_T(\theta_0) - G(\theta_0)| \\
& \leq B_T h[d(\hat{\theta}_T, \theta_0)] + | G_T(\theta_0) - G(\theta_0)|.
\end{array}
$
\medskip

\nin We know that if $X_T \stackrel{\rm P}{\rm \rightarrow}0$, $Y_T \stackrel{\rm P}{\rm \rightarrow}0$ $=> X_T + Y_T \stackrel{\rm P}{\rm \rightarrow}0$, i.e. the sum of $o_p(1)$ is $o_p(1)$.

\nin Under condition iii) $|G_T(\theta_0) - G(\theta_0)| = o_p(1)$. It remains to be shown that $B_T h[d( \hat{\theta}_T, \theta_0)]$ is $o_p(1)$.

\nin We have:
$$[B_T < M \; \mbox{and} \; h[d( \hat{\theta}_T, \theta_0)] < \epsilon/M] => [B_T h[d( \hat{\theta}_T, \theta_0)]] < \epsilon]$$
$$ \iff \left( ( B_T  < M) \cap  [h[d( \hat{\theta}_T, \theta_0)]
< \epsilon/M] \right) \subset [ B_T h[d( \hat{\theta}_T, \theta_0)] < \epsilon]$$

\nin Consider the complement: $(A \cap B)^c = A^c \cup B^c$. We get:

$((B_T >  M) \cup ( h[d( \hat{\theta}_T, \theta_0)]
> \epsilon/M) \supset [ B_T h[d( \hat{\theta}_T, \theta_0)]
> \sqrt{\epsilon}]$

\medskip
\nin It follows that:

\medskip
$ 
\begin{array}{ll}
P[B_T h[d( \hat{\theta}_T, \theta_0)]
> \epsilon] & \leq P [ (B_T >  M) \; \cup \; (h[d( \hat{\theta}_T, \theta_0)]
> \epsilon/M)] \\ 
 & \leq P [ B_T >  M] + P [h[d( \hat{\theta}_T, \theta_0)]
>h^{-1}(\epsilon/M)],
\end{array}
$

\nin because $P[A \cup B] \leq P(A) + P(B)$.

\nin Then, for any $\epsilon$, we can choose a value of $M$ and a number of observations $T$ sufficiently large to get
$P[B_T h[d( \hat{\theta}_T, \theta_0)] > \epsilon]$ arbitrarily small. Therefore, $B_T h[d( \hat{\theta}_T, \theta_0)]$ tends to zero in probability.
\hfill QED

\medskip
Then, the lemma can be applied with $G_T(\theta) = \frac{1}{T} \sum_{t=2}^T g(x_t, x_{t-1}; \theta)$ and $g(x_t, x_{t-1}; \theta) =  \ln | \phi + \frac{x_t - \phi x_{t-1}}{\sqrt{\omega + \alpha x_{t-1}^2}} \sqrt{\alpha}| $.
Under assumptions A.1, A.4, A.7, conditions i), ii), iii) are satisfied. For example:

$$G_T(\theta) \stackrel{\rm P}{\rm \rightarrow} E_0 g(x_t, x_{t-1}; \theta),$$

\nin by the weak law of large numbers applied to the transformation $g(x_t, x_{t-1}; \theta)$ of the ergodic stationary process $(x_t)$. Assumption A.8 corresponds to condition iv) of the lemma.

\subsection{Proof of Proposition 4}

a) Proof of convergence

We need to show that 
$ \hat{\phi}_T^2 + \hat{\alpha}_T \rightarrow  \phi_0^2 + \alpha_0$ 
in probability if
$(\hat{\phi}, \hat{\alpha}) \rightarrow (\phi_0, \alpha_0)$ in probability when $T \rightarrow  \infty$.

\nin Thi is a consequence of the "in probability" version of the continuous mapping theorem given in Appendix A.2


\medskip

b) Proof of Normality

\nin The Taylor series expansion pre-multiplied by $\sqrt{T}$ implies:

$ \sqrt{T}  \left[ (\hat{\phi}_T^2 + \hat{\alpha}_T) -  (\phi_0^2 + \alpha_0) \right]= \
 \left( \begin{array}{c} 2 \phi_0 \\ 1 \end{array} \right)'  \sqrt{T} \left( \begin{array}{c}\hat{\phi}_T - \phi_0 \\ \hat{\alpha}_T - \alpha_0 \end{array} \right) + o_p(1)$
 
$ = A' \sqrt{T}  \left( \begin{array}{c}\hat{\phi}_T - \phi_0 \\ \hat{\alpha}_T - \alpha_0 \end{array} \right) + o_p(1)$

\nin where $A'= [2 \phi_0 \;\; 1]$. We get the asymptotic normal distribution of $\hat{\xi}_T$:

$$\sqrt{T} (\hat{\xi}_{T} - \xi_0) \sim N( 0, V_{\xi}),$$
 
\nin where $V_{\xi} =   A' \Omega^* A$. The matrix $\Omega^*$ is:
$\Omega^* = diag(\Sigma^{-1}, V(\hat{\alpha}))$ where $\Sigma = E_0 (y^2/(\omega_0 + \alpha_0 y^2)$ given in Section 4.3.1. Matrix $V(\hat{\alpha})=(E_0 \frac{1}{(\omega_0 + \alpha_0 y^2)^2})/\tilde{V}_0(y^2)$ and $\tilde{V}_0(y^2) = \tilde{E}_0(y^4) - (\tilde{E}_0 y^2)^2$. In this formula, $\tilde{E}_0$ denotes the expectation of variables $y^4 \frac{1}{(\omega_0 + \alpha_0 y^2)^2}/E_0 \frac{1}{(\omega_0 + \alpha_0 y^2)^2}$ and $y^2 \frac{1}{(\omega_0 + \alpha_0 y^2)^2}/E_0 \frac{1}{(\omega_0 + \alpha_0 y^2)^2}$ [see, section 4.3.1 and Ling (2004) for the variance estimator formula].

\section*{Appendix B:  Simulation Results}

The purpose of this section is to illustrate the derived results in Appendix A using simulation experiments. We distinguish the case where the distribution of the innovation $\eta$ is known and the case it is not. 

First, we use the result in Proposition 1 and plot the Lyapunov exponent $\lambda=E(ln(|\phi+\sqrt{\alpha}\eta|))$ for different values of the parameter $\phi$ and $\alpha$. To do so, we assume $\eta\sim U[-1,1]$. Figure \ref{surface} shows that the Lyapunov exponent $\lambda$ remains lower than zero as as $\phi$ varies in $\{-1, -0.8, -0.6, \ldots, 0.6, 0.8, 1\}$ and $\alpha$ varies in $ \{0,0.1,0. 2, 0.3, \ldots, 0.8, 0.9,1\}$. 

Second, we assume $\eta\sim N(0,1)$, set the true parameters to $\phi_0=0.7$,  $\alpha_0=0.5$,  $\omega_0=0.01$. Note that the parameters are chosen to be close to their estimated value from the entire data in our application. The estimated densities are based on $4,000$ simulations and obtained via kernel density estimation. 
 Figure \ref{measures} plots the estimated density for the Lyapunov exponent $\hat\lambda_{2,T}$  and the stability measure $\hat\xi_{T}$ when the three parameters are estimated from a sample of size $T$ and plugged in. The results on panel (a) of the figure show that the 
mostly frequent estimated value is below zero for  $T=50$ or $T=100$, implying valid inference. In addition, panel (b) of Figure \ref{measures} shows that the density of the estimated alternative stability measure $\hat\xi_{T}=\hat\phi^2_T+\hat\alpha_T$ has its mode around the true value of $\xi$, which is $\xi_{0}=\phi_0^2+\alpha_0=0.99$. 

As a by-product, we present Figure \ref{estimates_v2}, which shows the estimated density for $\hat\phi_T$, $\hat\alpha_T$ and $\hat\omega_T$ for $T=50$ and $T=100$. The three panels in the figure provide evidence that the three parameters are fairly accurately estimated. More specifically, the estimated values have modes close to their true unknown parameters. The accuracy improves as the sample size increases from $T=50$ to $T=100$. 

\renewcommand\thefigure{B\arabic{figure}}  

\setcounter{figure}{0}

\begin{figure}[h!] \centering
\includegraphics[width=\linewidth,height=10cm]{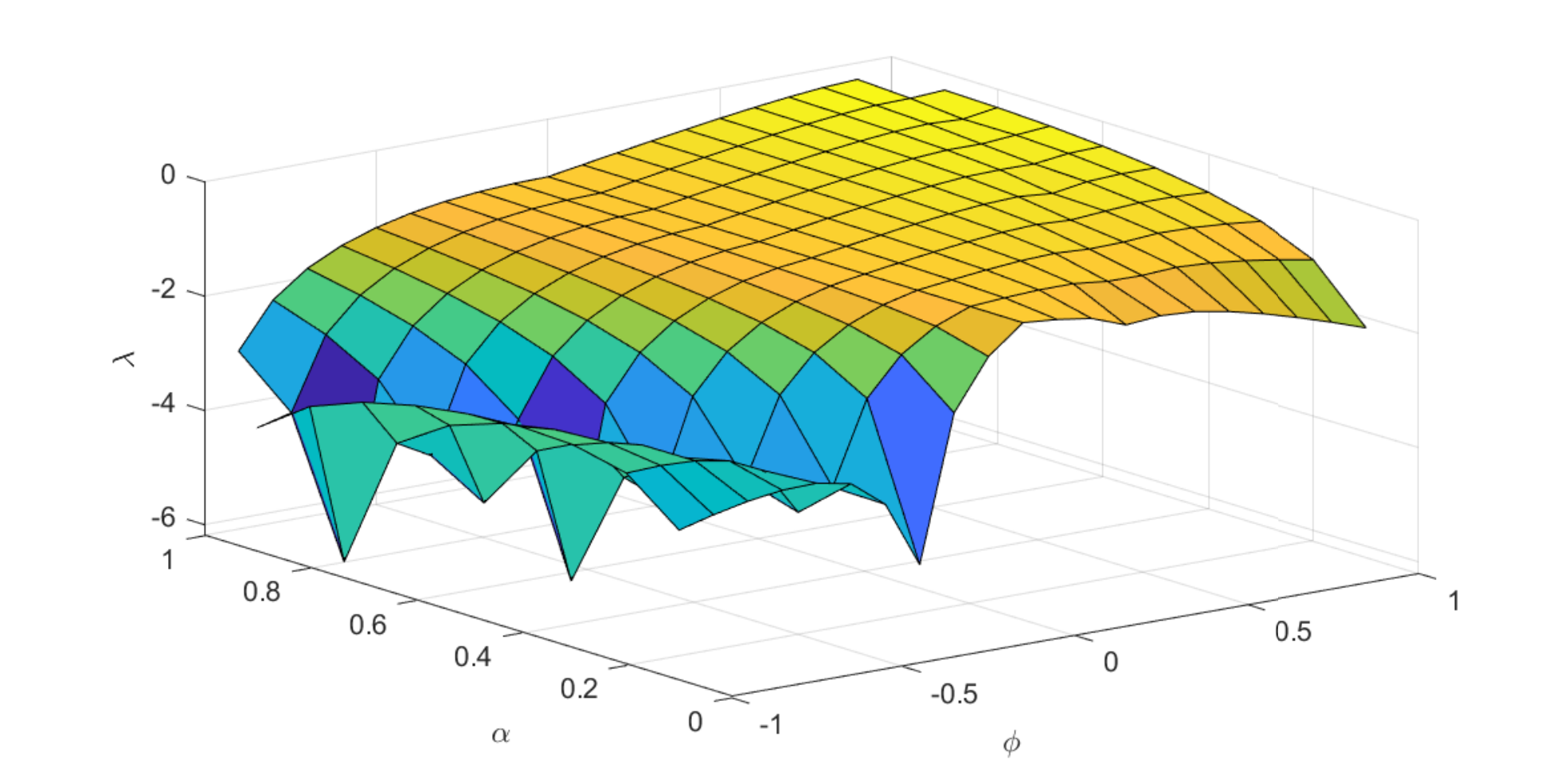} \centering 
\caption{Lyapunov exponent $\lambda=E(ln(|\phi+\sqrt{\alpha}\eta|))$ in terms of $\phi$ and $\alpha$ when $\eta\sim U[-1,1]$} 
\label{surface}
\end{figure}

\begin{center} 
\begin{figure}[H] 

\centering
\begin{subfigure}[b]{\linewidth} \includegraphics[width=\linewidth,height=8cm]{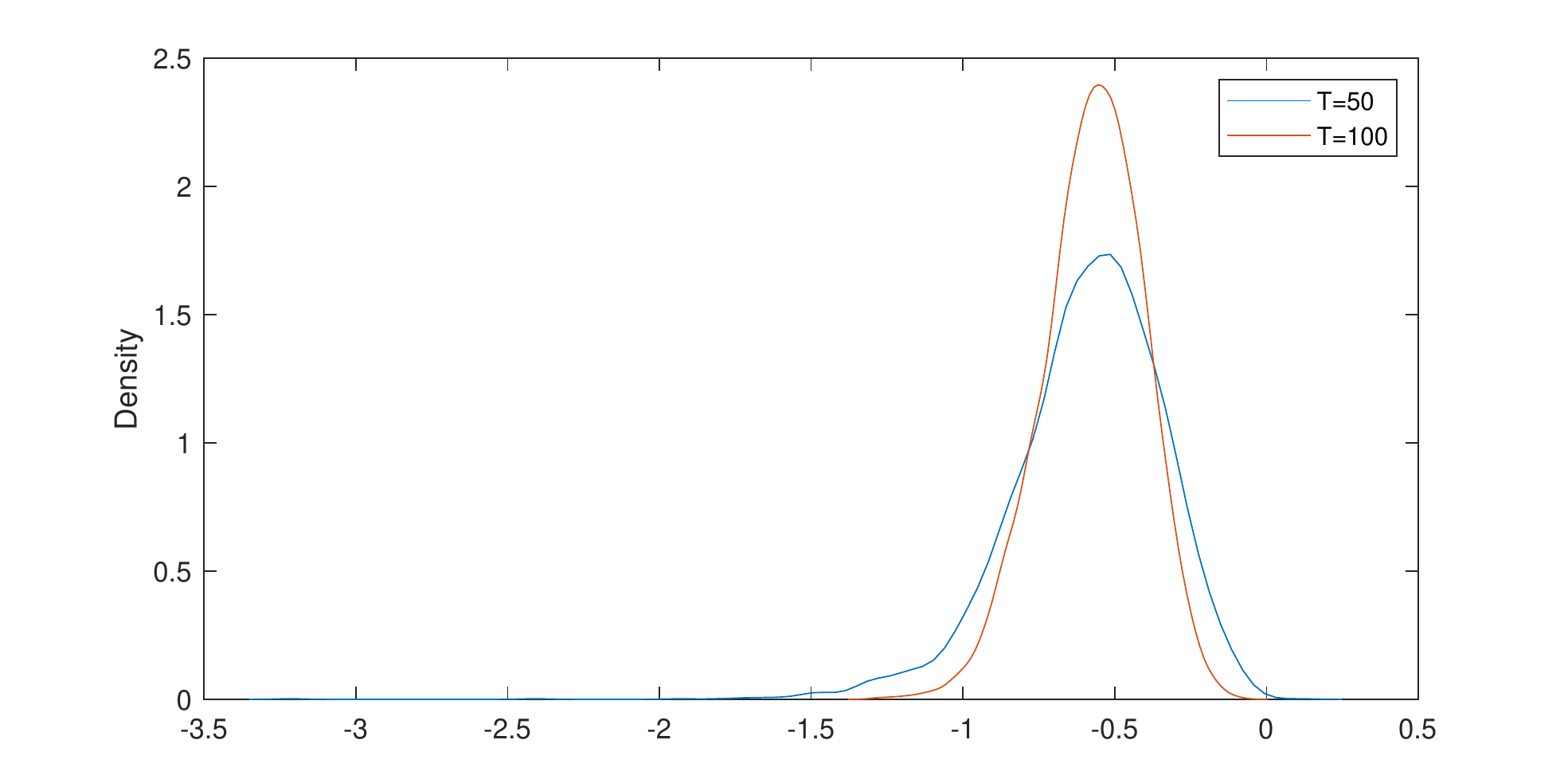} \centering \caption{Density of the Lyapunov exponent $\hat\lambda_{2,T}$} \end{subfigure}

\medskip

\begin{subfigure}[b]{\linewidth} \includegraphics[width=\linewidth,height=8cm]{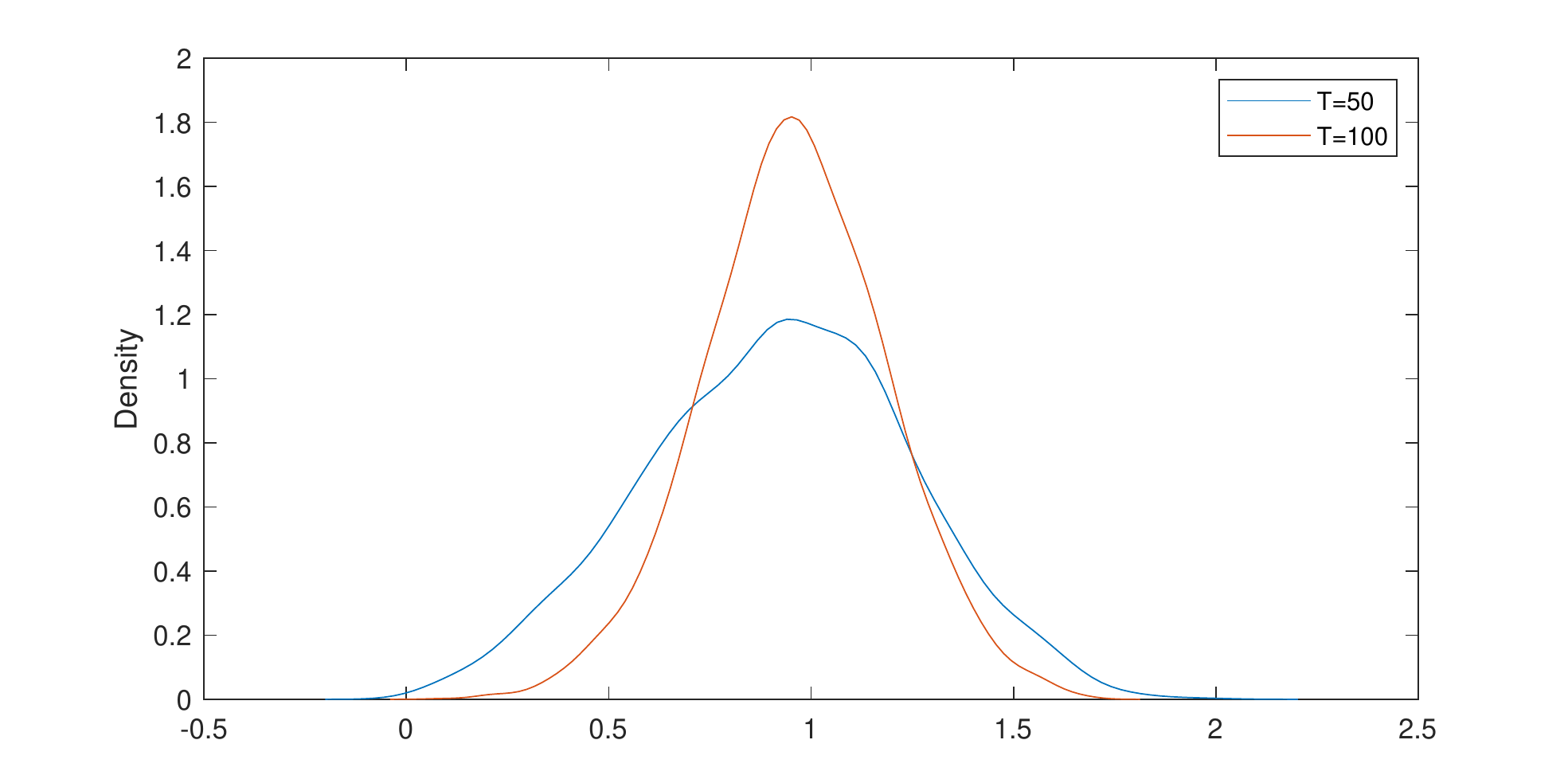} \centering \caption{ Density of the alternative stability measure $\hat\xi_{T}$} \end{subfigure}

\caption{Densities for stability measures based on estimated parameters} 
\label{measures}
\end{figure} 

\end{center}

\clearpage
\begin{center} 
\begin{figure}[H] \centering 
\begin{subfigure}[b]{\linewidth} \includegraphics[width=\linewidth,height=6cm]{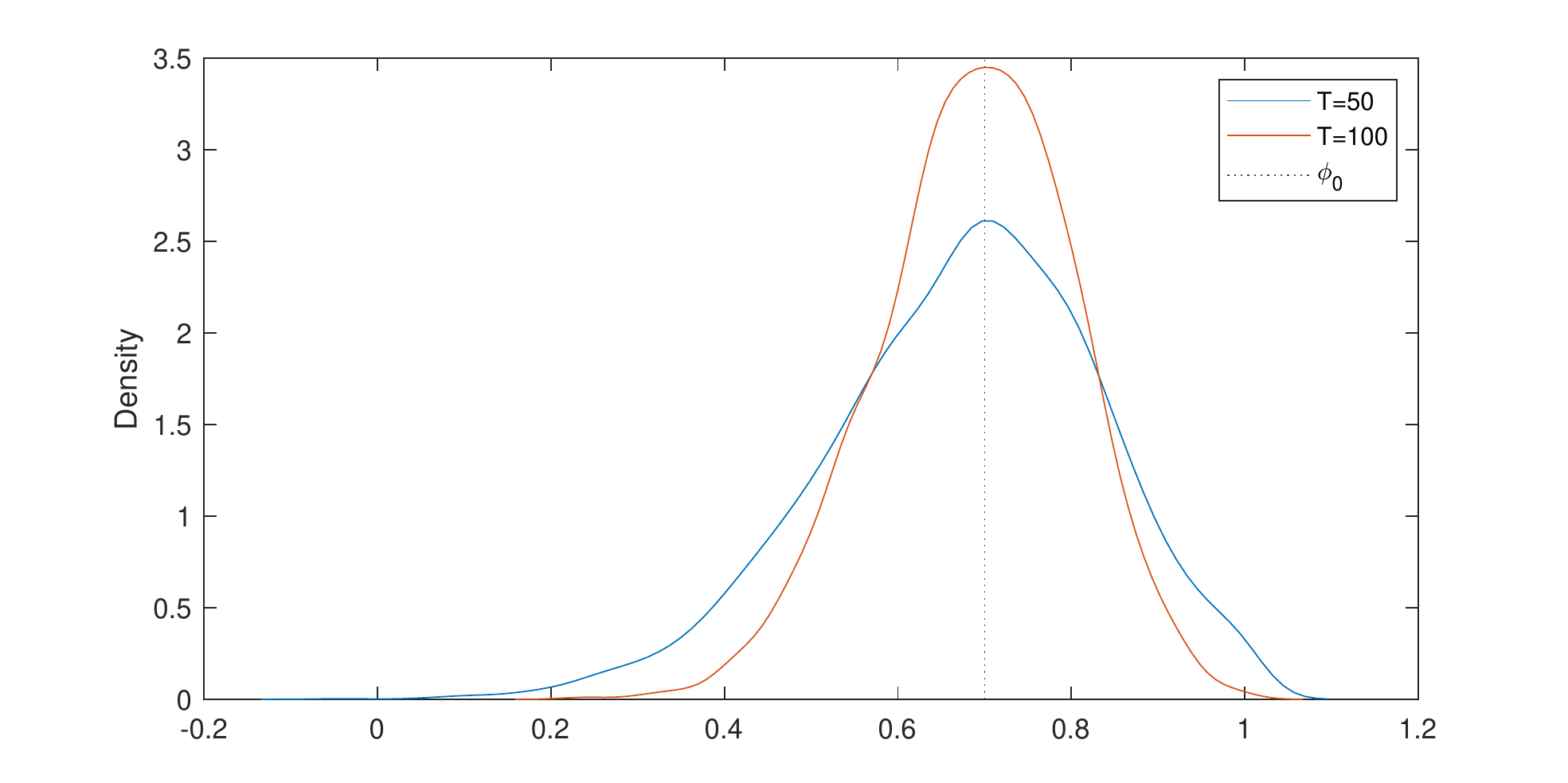} 
\centering 
{\caption*{(a)  Density of $\hat\phi_T$}}
\end{subfigure}
\begin{subfigure}[b]{\linewidth} \includegraphics[width=\linewidth,height=6cm]{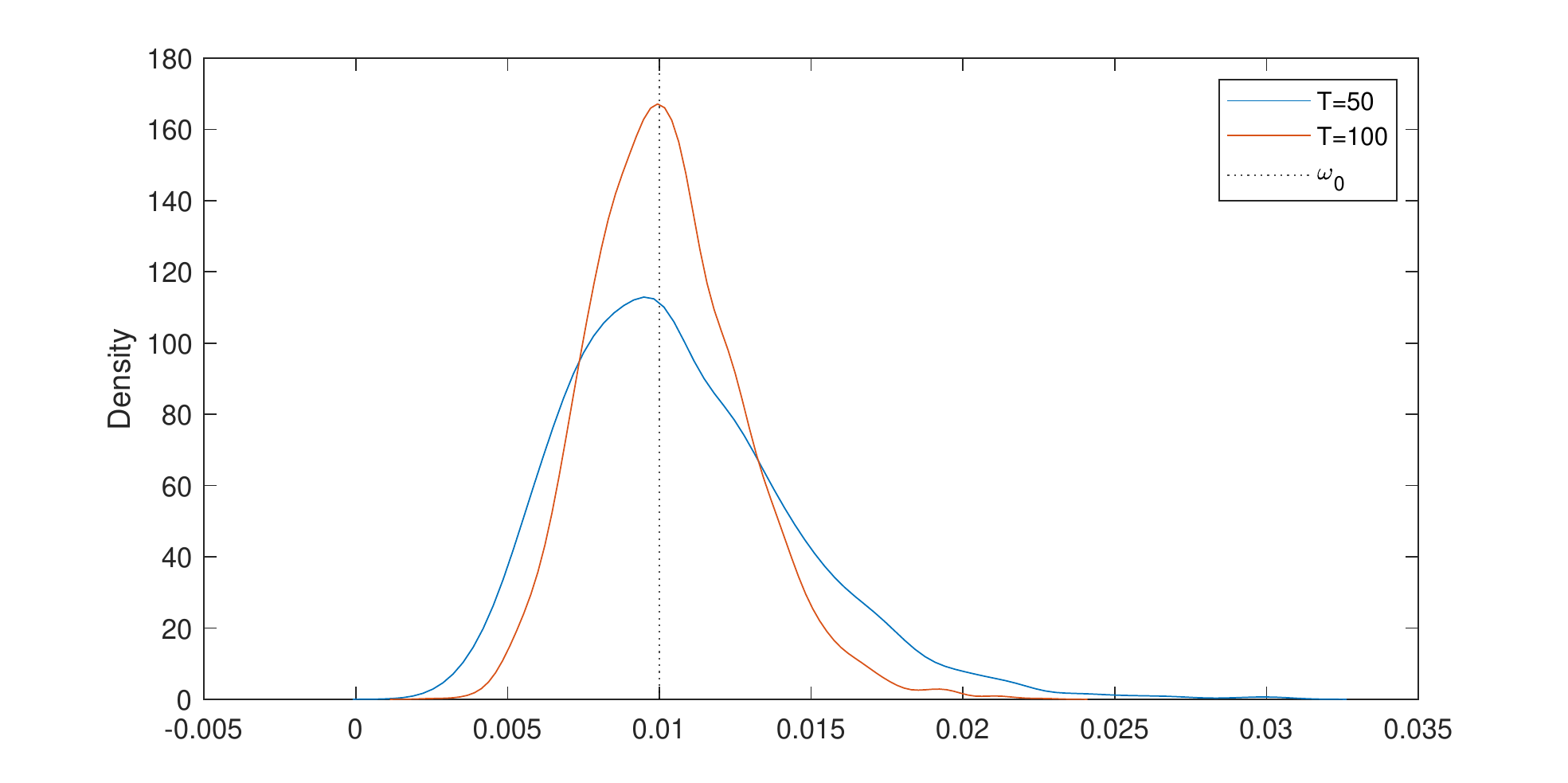} 
\centering 
 {\caption*{(b) Density of $\hat\omega_T$}}
 \end{subfigure}
\begin{subfigure}[b]{\linewidth} \includegraphics[width=\linewidth,height=6cm]{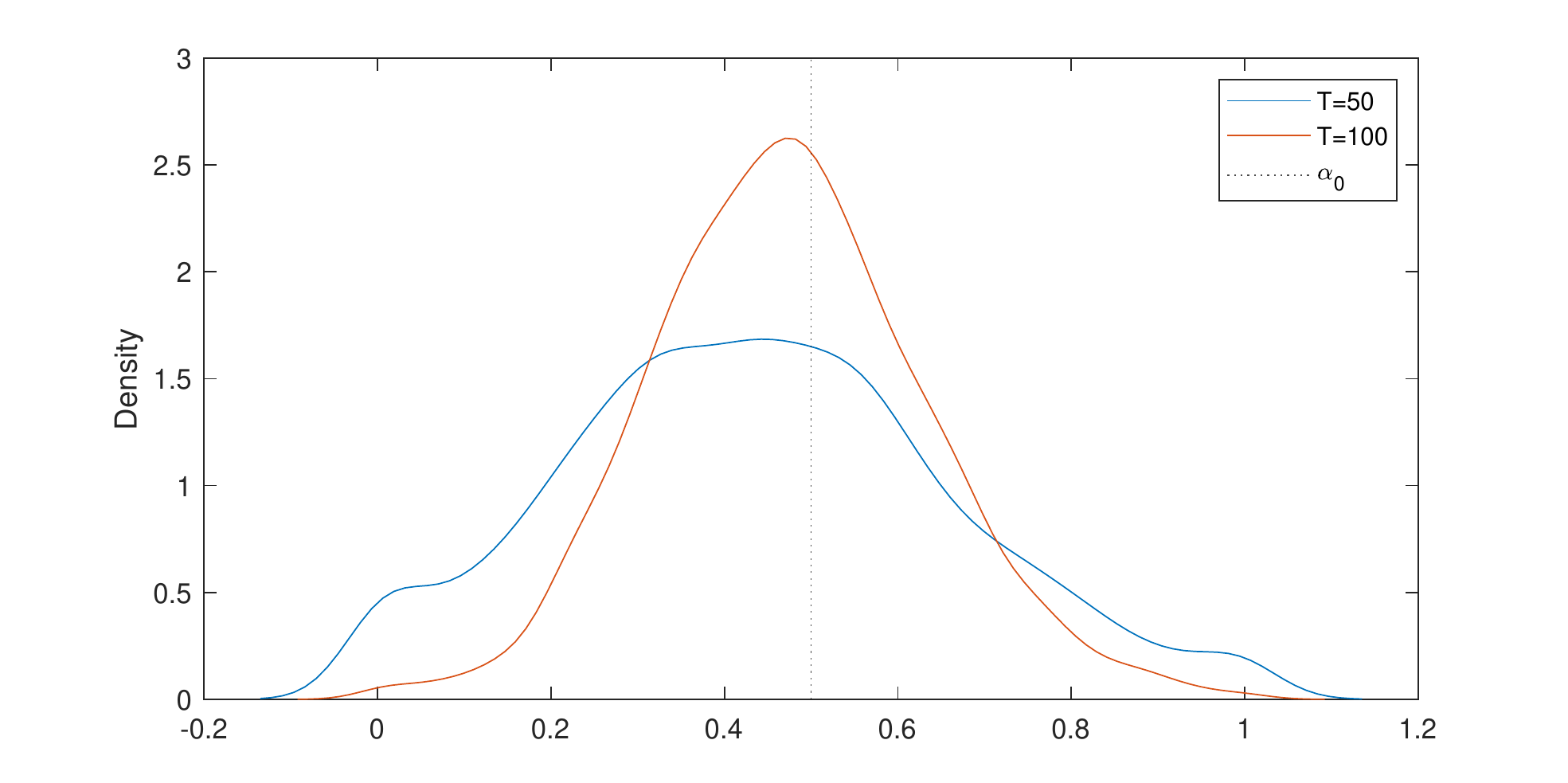} 
\centering 
{\caption*{(c) Density of $\hat\alpha_T$}}
 \end{subfigure}
 \caption{Densities of estimators for $\phi$, $\alpha$ and $\omega$} 
\label{estimates_v2}
\end{figure} 
\end{center} 

\section*{Appendix C:  More Empirical Results}

\renewcommand\thefigure{C\arabic{figure}}  

\setcounter{figure}{0}

\begin{figure}[H]
\centering
  \centering
  \includegraphics[width=15cm,height=15cm,angle = 0]{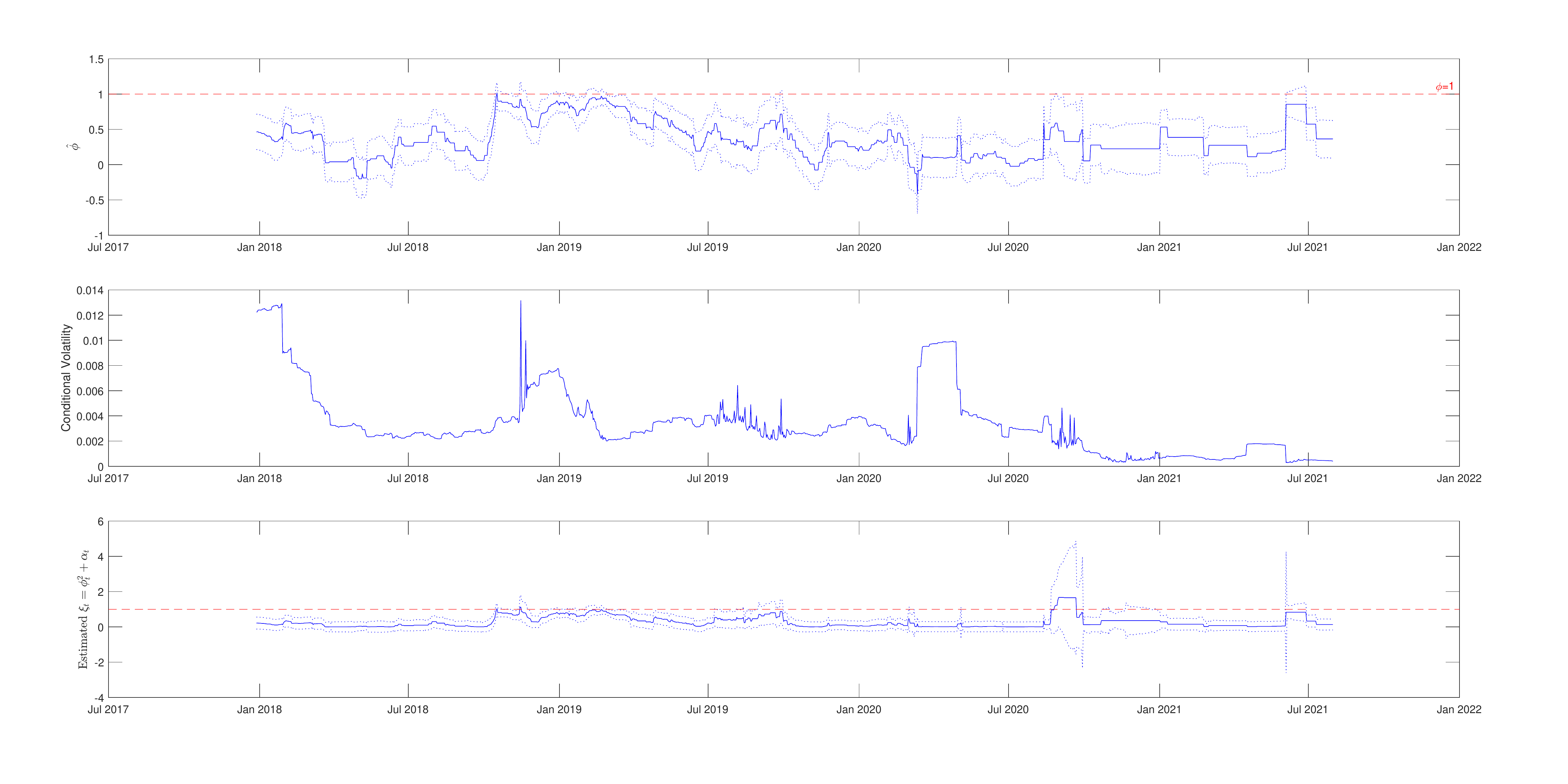}
\caption{tvDAR(1) parameter $\phi(t/T)$ and Lyapunov exponent $\xi(t/T)$}
 \label{DAR3}
  \end{figure}

Given that the proposed stability measure can be used as a mechanical tool to detect periods of instability in stablecoins, we use the results of Proposition 4 in Appendix A to construct an interval for $\xi$ employing the same rolling window approach as before. Figure \ref{DAR3} presents the results and contains, in its first panel, the estimated coefficient DAR model using the rolling windows approach, in its second panel, the conditional heteroskedasticity, and in the third panel, the Lyapunov exponent over time. In addition to the episodes of high persistence mentioned above, we observed, around September 2020, an important instability that is not due to high persistence in Tether price, but more frequent changes in the conditional heteroskedasticity, which can be seen in the second panel. 
This period can also be linked to higher local volatility in the observed data in Figure \ref{events}. There is no specific event we can associate with this movement, as is sometimes the case in crypto markets. However, the proposed model allows capturing those changes. 

As explained above, the measure of stability $\xi$ plotted in Figure \ref{DAR3} is more conservative than the Lyapunov exponent $\lambda$. Because we can have  $\xi\geq1$ while $\lambda<0$ so that valid inference is still possible, the rejection of  $\xi<1$ should be interpreted as the need for investors, regulators or stablecoin issuers to be cautious when predicting Tether future price around the tested periods.

\newpage

\begin{center}
\textbf{\Large{References}}
\end{center}
\nin Allen, F., Gu, X., and J. Jagtiani (2022): ``Fintech, Cryptocurrencies, and CBDC: Financial Structural Transformation in China'', Journal of International Money and Finance, Elsevier, vol. 124(C).
\medskip

\nin Andrews, D. (1987): ``Consistency in Nonlinear Econometric Models: A Generic Uniform Law of Large Numbers'', Econometrica, 55, 1465-1471.
\medskip

\nin Barry, C. B., and R. L., Winkler (1976) : ``Nonstationarity and Portfolio Choice'', The Journal of Financial and Quantitative Analysis, Vol. 11, No. 2, pp. 217-235.

\medskip

\nin Bandi, F., and P., Phillips (2009) : ``Nonstationary Continuous-Time Processes'', in Handbook of Financial Econometrics, Y., Ait Sahalia, and L., Hansen eds., 140-199, Elsevier.\medskip

\nin Baum\"ohl, E. and T. Vyrost (2020) : ``Stablecoins as a crypto safe haven? Not all of them!'', ZBW-Leibniz Information Centre for Economics, Kiel, Hamburg
\medskip

\nin Bianchi, D., Rossini, L and M., Iacopini (2022) ``Stablecoins and Cryptocurrency Returns: What is the Role of Tether'',
Working Paper, University of Milan?
\medskip

\nin Billingsley, P. (1999): ``Convergence of Probability Measures'', New York, Wiley.
\medskip

\nin Blanchard, O. and M. Watson (1982): ``Bubbles, Rational Expectations, and Financial Markets'', P. Wachtel (ed.) Crisis in the Economic and Financial Structure, Lexington Books, Lexington, Mass.
\medskip

\nin Borkovec, M. (2000): ``Extremal Behavior of the Autoregressive Process with ARCH(1) Errors'', Stochastic Processes and their Applications, 85, 189-207.
\medskip

\nin Borkovec, M. and C. Kluppenberg (2001): ``The Tail of the Stationary Distribution of the Autoregressive Process with ARCH(1) Errors'', Annals of Applied Probability, 11, 1220-1241.
\medskip

\nin Bullman, D., Klemm, J., and A.,Pinna (2019): ``In Search for Stability in Crypto-assets: Are Stablecoins the Solution?'', European Central Bank Occasional Paper Series No 230
\medskip

\nin Catalini, C., and A.,de Gortari (2021): ``On the Economic Design of Stablecoins'', Available at SSRN: https://ssrn.com/abstract=3899499 or http://dx.doi.org/10.2139/ssrn.3899499.\medskip

\nin Chen, M., Li, D. and S. Ling (2014): ``Non-Stationarity and Quasi-Maximum Likelihood Estimation on a Double Autoregressive Model'', 
Journal of Time Series Analysis, 35: 189--202. 
\medskip

\nin Chen, M., Qin, C., and X., Zhang (2022): ``Cryptocurrency price discrepancies under uncertainty: Evidence from COVID-19 and lockdown nexus,'' Journal of International Money and Finance, Elsevier, vol. 124(C).
\medskip

 \nin Chandler, G., and W., Polonik (2012):  ``Mode Identification of Volatility in Time-Varying Autoregression'', Journal of the American Statistical Association, 107(499), 1217-1229.

\medskip

\nin Chandler, G., and W., Polonik (2017): `` Residual Empirical Processes and Weighted Sums for Time-Varying Processes with Applications to Testing for Homoscedasticity'', Journal of Time Series Analysis, vol. 38, 72-98.

\medskip

\nin Dahlhaus, R. (2000): ``A Likelihood Approximation for Locally Stationary Processes'', Annals of Statistics, 28, 1782-1794. \medskip

\medskip

\nin Dahlhaus, R., S. Richter and W. Wu (2019): ``Towards a General Theory for Nonlinear Locally Stationary Processes'',
Bernoulli, 25, 1013-1044.\medskip

\medskip

\nin Day, W. (1976): ``A Reform of the European Currency Snake'', IMF Econ Rev 23, 580?597.

\medskip

\nin Dechert, W.D. and  R. Gencay (1992): ``Lyapunov Exponents as a Nonparametric Diagnostic for Stability Analysis'', Journal of Applied Econometrics, VOL. 7, S41-S60 
\medskip

\nin Fan, J., M. Farmen and I. Gijbels (1998): ``Local Maximum Likelihood Estimation and Inference'', Journal of the Royal Statistical Society, series B, 60, Part 3, 591-608.
\medskip

\nin Froot, K., and M. Obstfeld (1991): ``Intrinsic Bubbles: The Case of Stock Prices'', American Economic Review, 81, pp. 1189-1214.
\medskip

\nin Gourieroux, C.: ``ARCH Models and Financial Applications'', New York: Springer-Verlag, 1997.
\medskip

\nin Gourieroux C. and J. Jasiak (2019): ``Robust Analysis of the Martingale Hypothesis'', Econometrics and Statistics, Vol 9, 17-41.
\medskip

\nin Gourieroux, C., and J.M., Zakoian (2017): ``Local Explosion Modelling by Noncausal Processes'', Journal of the Royal Statistical Society (JRSS), Series B, 79, 737-756.
\medskip

\nin Griffin, J. M. , and A.,Shams (2020): ``Is Bitcoin Really Untethered?'',  The Journal of Finance, vol. 75, issue 4
\medskip

\nin Hong, H., and J. C., Stein (2002): ``A Unified Theory of Underreaction, Momentum Trading, and Overreaction in Asset Markets'', The Journal of Finance, 54, issue 6, 2143-2184

\medskip

\nin Huisman, R., Koedijik, K.G., and Pownall, R.A.J., (1998): ``VaR-x: Fat Tails in Financial Risk Management'',
Papers 98-54, Southern California - School of Business Administration.
\medskip

\nin Kortian, T. (1995): ``Modern Approaches to Asset Price Formation: A Survey of Recent Theoretical Literature'',
RBA Research Discussion Papers rdp9501, Reserve Bank of Australia.
\medskip

\nin Lebaron, B. (1994): ``Chaos and Nonlinear Forecastability in Economics and Finance'', Philosophical Transactions of the Royal Society of London. Series A: Physical and Engineering Sciences, 348, 397-404.
\medskip

\nin Li, Q. (1999): ``Consistent Model Specification Tests for Time Series  Econometric Models'', Journal of Econometrics, 92, 101-147.
\medskip

\nin Li, W. K. (1992): ``On the Asymptotic Standard Errors of Residual Autocorrelations in Nonlinear Time Series Modeling'',
Biometrika, 79, 435-437.
\medskip

\nin Li, W. K. and Mak, T. K. (1994): ``On the Squared Residual Autocorrelations in Non-linear Time Series with Conditional
Heteroskedasticity'', Journal of Time Series Analysis, 15, 627-636.
\medskip

\nin Li, D., Guo, S., and K., Zhu (2019): ``A Double AR Model without Intercept: An Alternative to Modeling Nonstationarity and Heteroscedasticity'',  Econometric Reviews, 38, issue 3, 319-331.
\medskip

\nin Li, D., Ling, S. and R., Zhang (2016): ``On a Threshold Double Autoregressive  Model'', Journal of Business and Economic Statistics, 34, 68-80.
\medskip

\nin Li, Y. and Mayer, S., (2022) ``Money Creation in Decentralized Finance: A Dynamic Model of Stablecoin and Crypto Shadow Banking'', Fisher College of Business Working Paper No. 2020-03-030, Charles A. Dice Center Working Paper No. 2020-30
\medskip

\nin Ling, S. (2004): ``Estimation and Testing Stationarity for Double Autoregressive Models'', JRSS Series B, 66,  63-78
\medskip

\nin Ling, S. (2007): ``A Double AR(p) Model: Structure and Estimation", Statistica Sinica'', Vol. 17, No. 1., 161-175
\medskip

\nin Ling, S. and D. Li (2008):  ``Asymptotic Inference for a Nonstationary Double AR(1) Model'', Biometrika , 95, 1, pp. 257–263
\medskip

\nin Liu, F, Li, D. and X. Kang (2018) ``Sample Path Properties of an Explosive Double Autoregressive Model'', Econometric Reviews, 37, 484-490.
\medskip

\nin Lyons, R. K., and  G., Viswanath-Natraj (2020): ``What Keeps Stablecions Stable?'', NBER Working Papers 27136, National Bureau of Economic Research, Inc.
\medskip

\nin Nelson, D.B. (1990): ``Stationarity and Persistence in the GARCH(1,1) Model'', Econometric Theory, 6, 318-334.
\medskip

\nin Newey, W. (1991): ``Uniform Convergence in Probability and Stochastic Continuity'', Econometrica, Vol 59, 1161-1167.
\medskip

\nin Potcher, B. and J. Prucha (1989): ``Uniform Law of Large Numbers for Dependent and Heteregeneous Processes,'' Econometrica, 57, 675-683.
\medskip

\nin President's Working Group (2021): ``President's Working Group on Financial Markets Releases Report and Recommendations on Stablecoins'', https://home.treasury.gov/news/press-releases/jy0454.
\medskip

\nin Sprott, J.C. (2003): ``Chaos and Time-Series Analysis'', Oxford University Press, Oxford 
\medskip 

\nin Sprott, J.C. (2014): ``Numerical Calculation of Largest Lyapunov Exponent'', working paper, University of Wisconsin.
\medskip

\nin Wang, G., Ma, X., and H., Wu. (2020): ``Are Stablecoins truly diversifiers, hedges, or Safe Havens against traditional cryptocurrencies as their names?'',  Research in International Business and Finance, 54, p. 101-225.
\medskip

\nin Zakoian, J.M. (1994): ``Threshold heteroskedastic models'',  Journal of Economic Dynamics and Control, 18, 931-955.

\end{document}